\documentclass[11pt]{amsart}
\usepackage[english]{babel}
\usepackage{inputenc}

\usepackage{latexsym}
\usepackage{amssymb}
\usepackage{amsmath}
\usepackage{amsfonts}
\usepackage{stmaryrd}
\usepackage[mathscr]{eucal}
\usepackage{xcolor}

\usepackage{amscd}

\usepackage{graphicx}
\usepackage{subcaption}
\usepackage{epsfig}
\usepackage{relsize}
\usepackage{fullpage}

\usepackage{longtable}
\usepackage{rotating}
\usepackage{array}
\usepackage{multicol}
\usepackage{multicoltab}
\usepackage{multirow}
\usepackage{supertabular}
\usepackage{makecell}
\usepackage{mathtools}
\usepackage{threeparttable}
\usepackage{booktabs}

\usepackage{amsthm}
\usepackage[all,cmtip]{xy}
\usepackage[pdfstartview=FitH]{hyperref}
\usepackage{comment}
\usepackage{psfrag}
\usepackage{verbatim}
\usepackage{hyperref}
\hypersetup{
colorlinks=true,
linkcolor=blue,
allcolors=blue
}

\usepackage{pifont}
\usepackage{caption}
\usepackage{csquotes}

\usepackage{soul}
 \usepackage[normalem]{ulem}

\usepackage{tikz}
\usetikzlibrary{positioning,fit,calc,intersections,through,backgrounds,perspective,arrows.meta}
\usepackage{pgfplots}
\pgfplotsset{compat=1.15}
\usepackage{mathrsfs}
\usetikzlibrary{arrows}

\definecolor{cstay}{HTML}{0072B2}
\definecolor{ccity}{HTML}{E69F00}
\definecolor{ccm}{HTML}{009E73}
\definecolor{cout}{HTML}{CC79A7}
\definecolor{ink}{HTML}{222222}
\definecolor{soft}{HTML}{F1F3F5}
\definecolor{rule}{HTML}{9CA3AF}

\usepackage{enumerate}
\usepackage{enumitem}

\usepackage[style=numeric,backend=biber,maxbibnames=99,sortcites=true]{biblatex}

\newtheorem*{thm*}{Theorem}
\newtheorem*{cor*}{Corollary}	

\theoremstyle{definition}
 
\newtheorem*{definition*}{Definition}    	

\numberwithin{equation}{section}

\begin{document}






















\title{Quantifying displacement: an urban expansion consequence via persistent homology}


\author[R.~Rodríguez]{Rita Rodríguez Vázquez$^{\text{1}}$}

\author[M.~Cuerno]{Manuel Cuerno$^{\text{2, A}}$}


\thanks{$^{\text{1}}$ Department of Quantitative Methods, CUNEF Universidad, Madrid, Spain. \texttt{rita.rodriguez@cunef.edu}}

\thanks{$^{\text{2}}$ Department of Mathematics, CUNEF Universidad, Madrid, Spain. \texttt{manuel.mellado@cunef.edu}}

\thanks{$^{\text{A}}$ M. Cuerno has been financially supported by the project ``Charting political ideological landscapes in Europe: Fault lines and opportunities (POL-AXES)'' - Programa Primas y Problemas 2023 from Fundación BBVA, and PID2021-124195NB-C32 and PID2024-158664NB-C22 
from the Ministerio de Econom\'ia y Competitividad de Espa\~{n}a (MINECO)}





\date{\today}



\begin{abstract}

Population displacement is a housing-related involuntary residential dislocation. It has become increasingly widespread in many cities, particularly in neighbourhoods undergoing rapid economic and demographic change, and measuring it is essential to assess the social consequences of urban transformation and housing market pressures. Despite its relevance, quantifying displacement presents difficulties due to limited replicability across cities and time periods and the need to analyse long time spans: displacement is a gradual process, impossible to capture in one data snapshot. We introduce a novel tool to overcome these difficulties. Using publicly available address change data, we construct four cubical complexes simultaneously incorporating geographical and temporal information of people moving, and analyse using Topological Data Analysis tools. Finally, we demonstrate this method through a 20-year case study in Madrid, Spain. 
The results demonstrate its ability to reveal patterns consistent with population displacement and identify the neighbourhoods and years potentially affected--patterns not observable from yearly maps of the raw address change data.
\end{abstract}

\setcounter{tocdepth}{1}

\maketitle



\section{Introduction}

Population displacement has been defined as a housing-related involuntary residential dislocation \cite{marcuse}.
The literature distinguishes several forms of population displacement linked to urban change. Marcuse \cite{marcuse} differentiates direct displacement, when residents are forced to move due to rent increases, eviction, or redevelopment, and exclusionary or indirect displacement, which occurs when rising housing costs prevent lower-income households from accessing a neighbourhood. Subsequent research also identifies chain displacement and cultural displacement, reflecting longer-term population replacement and changes in neighbourhood social dynamics \cite{freeman2005,newmanwyly,zuk,glass1964london,somashekhar}.
The widespread and uneven effect of displacement across social groups has motivated a broad range of studies, focusing on everything from characterising displaced individuals to identifying the potential causes and consequences of the phenomenon.

A wide range of approaches has been employed to assess the extent of displacement.
Survey-supported research detects displacement by explicitly inquiring about individuals’ motivations for relocation.
The seminal 1981 study conducted by the National Institute of Advanced Studies \cite{displacement1981single} examined who and why was moving out of the rapidly uplifting neighbourhood of Hayes Valley in San Francisco.
Researchers found that about one fourth of the movers between 1975-1979 left involuntarily, and disproportionately Black, eldery, or low-income residents.
More recent studies \cite{desmond15, desmond17} analyse the prevalence and characteristics of displacement using Milwaukee Area Renters Study survey data.
Of special relevance is the Panel Study of Income Dynamics (PSID), a longitudinal household survey conducted in the United States since 1968. By following the same individuals and families over time and recording their residential mobility, the PSID enables longitudinal analyses of neighbourhood change and has been employed in several studies investigating displacement \cite{freeman2005, ding, freeman24}.
The elevated costs and limited availability of survey data have encouraged alternative methods that infer displacement indirectly through socio-economic indicators, either by contrasting these measures between in-movers and out-movers or interpreting spikes as evidence of displacement. 
For instance, \cite{Ellen} investigates exit rates of low-income residents in neighbourhoods with increasing income, while McKinnish et al. \cite{mckinnish} focus on exit rates among vulnerable groups. 
Other studies compare a metrics's value in a neighbourhood with that in a control group, often aimed at understanding its relation with gentrification processes: Ding et al. \cite{ding} contrast mobility rates and destination outcomes between gentrifying and non-gentrifying tracts in Philadelphia, and Ellen et al. \cite{ellentorrats} compare demographic changes in gentrifying tracts with those in the metropolitan area to assess whether observed shifts indicate displacement or citywide dynamics. Finally, composite indices such as the Los Angeles Index of Displacement Pressure \cite{LA} weigh individual, household, and neighbourhood indicators to assess relative risk.

These approaches present a series of shortcomings, which fall into two categories. The first and perhaps the most important one is the lack of replicability to other cities and time periods, which is due to ad-hoc methodologies and data. Specifically, the choice of a metric and a control group lead to different definitions thereof, hindering the comparison of displacement rates, and surveys and specific indicators are tailored to the specific cities they are designed for, such as the Los Angeles Index of Displacement Pressure. Second, the time-scale of some analyses is too short to capture the full displacement processes, as it is not an immediate phenomenon but one that unfolds over time.
For instance, not all leases expire at the same time, redevelopment policies come into force progressively, and infrastructure and development projects such as transport or urban renewal are rolled out in phases.
This limitation, although frequent, does not apply to studies using longitudinal datasets such as the aforementioned PSID.

We propose a novel methodology to overcome these challenges and validate it in a case study for the city of Madrid, Spain, using solely address change data. Using this data, we build a suitable topological space and apply persistent homology, a key tool in Topological Data Analysis (TDA), to extract quantitative patterns potentially indicative of population displacement.
This data is not tied to any city-specific features and is also publicly available in many European countries. It may be found for the following countries; Denmark, Sweden, Norway, Finland, the Netherlands, Belgium, Austria, and Switzerland. Population registers record every residential move, allowing the national statistics authorities to construct origin–destination matrices of address changes between small areas within cities.
In the U.S., neighbourhood-level migration data coverage varies by city, and typically exists only through city administrative records or special datasets. Important such datasets include the Urban and Regional Migration Estimates \cite{migration_estimates}, which provides migration estimates for the major U.S. metro areas at neighbourhood level, and 
MIGRATE \cite{MIGRATE}, a dataset of annual migration matrices at census block level.
The broad availability of this data comes at a cost, however, as the reason behind each individual move remains unknown.

To the best of our knowledge, this is the first application of TDA techniques to population displacement.
Nevertheless, related literature employs persistent homology to study urban factors potentially linked to this phenomenon. From the standpoint of resource access, Hickok et al. \cite{Hickok_siam} developed a methodology to analyse access to polling sites in several major U.S. cities, while O’Neil and Tymochko \cite{tymochko_cooling} conducted a similar study focused on access to cooling centres; Tymochko is also currently extending the methodology of \cite{Hickok_siam} to the case of urban parks. More politically oriented applications include the work of Friesen and Ziegelmeier \cite{friesen_segregation} on racial segregation in the U.S., and Duchin et al. and Shah \cite{duchin_gerrymandering, shah_gerrymandering} on gerrymandering. From a methodological perspective, other works--such as \cite{erik_trigo}--have used 3D solid bodies constructed from cubical complexes for TDA purposes, but without incorporating time as one of the axes. 
Therefore, the present work is novel both in its content and in its methodological approach.

This work opens the door to applying the same methodology to investigate displacement in cities around the world. The data we use is collected in most countries, making our approach highly replicable. Taken together, these features address the two main critiques identified in the existing literature: the reliance on ad-hoc techniques and data, and inadequate time-scale.
While our data captures every residential move, it does not the underlying reasons for each one. Consequently, rather than implying that every recorded move constitutes displacement,
our analysis provides strong evidence suggestive of population displacement pattern.
The use of persistent homology makes our contribution both novel within the TDA framework and well grounded, as it relies on a construction tailored to this problem that fully exploits the strengths of persistent homology.

\section{Materials and methods}\label{sec:methods}

\subsection{Data}\label{sec:data}

\begin{figure}[tbhp]
\centering
\includegraphics[width=300pt]{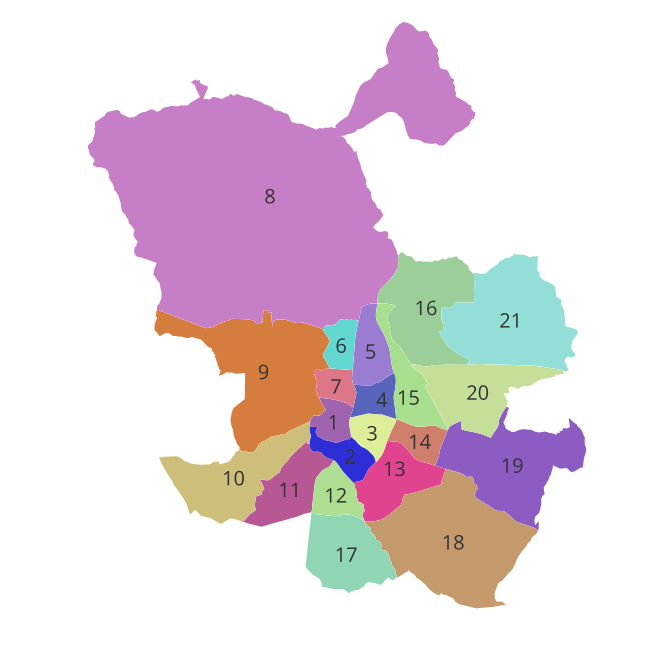}
\caption{Madrid map and their 21 districts as of April 2025. Base layer: Ayuntamiento de Madrid, ``Distritos municipales de Madrid''
(\url{https://datos.madrid.es/dataset/300497-0-distritos-municipales-madrid}), CC BY 4.0
(\url{https://creativecommons.org/licenses/by/4.0/}). Map created by the authors.}\label{fig:districts_num}
\end{figure}

The city of Madrid is organised into multiple administrative levels. It comprises 21 districts numbered 1 to 21 (see Figure \ref{fig:districts_num}), each subdivided into several neighbourhoods. 
Each one is uniquely identified by a numerical code in which the final digit denotes the neighbourhood and the preceding digits indicate the district.
For example, 26 designates the 6th neighbourhood in district 2, and 141 denotes the 1st neighbourhood in district 14.
In order to incorporate the spatial structure of these neighbourhoods into our study, we use a shapefile of the city neighbourhoods, available at \cite{geoportalmadrid}.

All the data used in this study is publicly available on the official website of the city council \cite{bancodedatosmadrid}. The main data sources are the tables in the `C. Demograf\'ia y Poblaci\'on' section, specifically those containing information about address changes with origin inside the city. This includes information on the number of residents who moved, their destination and the number of people who left the city. 
We use the finest spatial and temporal granularity available, which is yearly data at neighbourhood level. Each year's data may be represented as an origin-destination matrix, whose rows and columns represent the possible origins and destinations, respectively, for residents who moved that year. Thus, rows correspond exactly to the city neighbourhoods, whereas the columns include all the neighbourhoods together with other possible destinations for people moving outside the city.
We consider the full time series available as of April 2025, i.e. data from 2004 until 2023.

\subsection{Persistent homology}{\label{sec:pershom}}

Topology is a key field in mathematics that studies and classifies geometric spaces in terms of their invariance under perturbations such as stretching, folding, or even twisting. In particular, Algebraic Topology is the branch that uses algebraic tools to study these objects. Homology is fundamental in Algebraic Topology: given a space $X$, its homology groups $H_n(X)$ encode information about its $n$-dimensional holes, such as connected components ($n=0$)---a set of locations that are all reachable from one another, so that you can move between them without leaving the group---, loops ($n=1$), voids ($n=2$), and the analogues for $n>2$. 

At the beginning of this century, topology began to play a key role in data analysis \cite{carlsson_topology,zomorodian_carlsson,zomorodian_computing}, giving rise to Topological Data Analysis (TDA). The central idea is to study the shape of data (understood as the shape of the dataset viewed as a point cloud) by, for example, analysing the prominence of its $n$-dimensional holes. \emph{Persistent homology} is the primary tool for this. The importance of such holes are tracked through a filtration, that is, a nested family of combinatorial structures  (vertices, edges, triangles, tetrahedra, and their higher-dimensional counterparts) built from point-cloud data that evolves as a parameter grows. As it increases, new $n$--dimensional tetrahedral structures are incorporated. At each step of the filtration, the homology groups are computed, and it is thus pertinent to study whether homology groups existing in previous steps of the nested sequence are still ``alive''. These groups or $n$--dimensional holes that appear and later disappear will also be referred as \textit{topological features} in the rest of the paper.

The resulting features are summarised in barcodes or persistence diagrams, where each interval $[b,d]$ (or point $(b,d)$) indicates that the $n$--dimensional hole appears at scale $b$ and vanishes at $d$, where $b\leq d$ as they indicate ordered times. Points near the diagonal, that is, $b$ close to $d$, correspond thus to short-lived structures; see, e.g., the illustrative examples in \cite{ghrist}. 
Persistence is defined as $b-d$, i.e. the length of the bars. Features with high persistence are generally considered strong topological signals, while those with low persistence are often regarded as noise.
For readers interested in formal definitions of homology, simplicial complexes, and related notions, we refer to \cite{hatcher}.

\subsection{Cubical complexes for grayscale images}\label{sec:cubical}

\begin{figure}[tbhp]
\centering
\includegraphics[width=.96\linewidth]{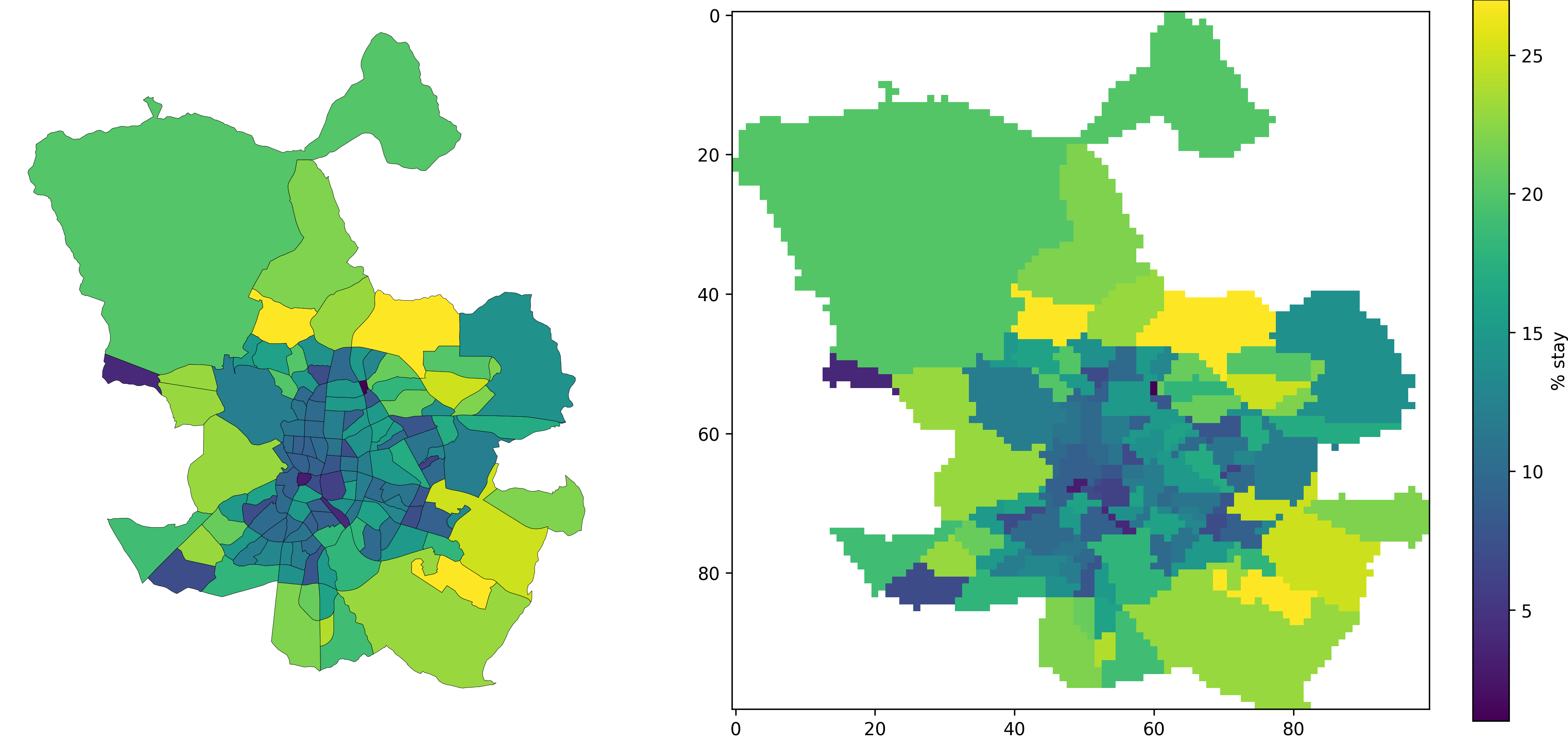}
\caption{Grayscale image depicting the percentage of people moving in 2023 that stayed in the same neighbourhood (left) and resulting $100 \times 100$ covering the city map (right). Base layer: Ayuntamiento de Madrid, ``Barrios municipales de Madrid''
(\url{https://datos.madrid.es/dataset/300496-0-barrios-madrid}), CC BY 4.0
(\url{https://creativecommons.org/licenses/by/4.0/}). Map created by the authors.}
\label{fig:gray_to_cubical}
\end{figure}

\begin{figure}[ht]
    \centering
    \includegraphics[width=\linewidth]{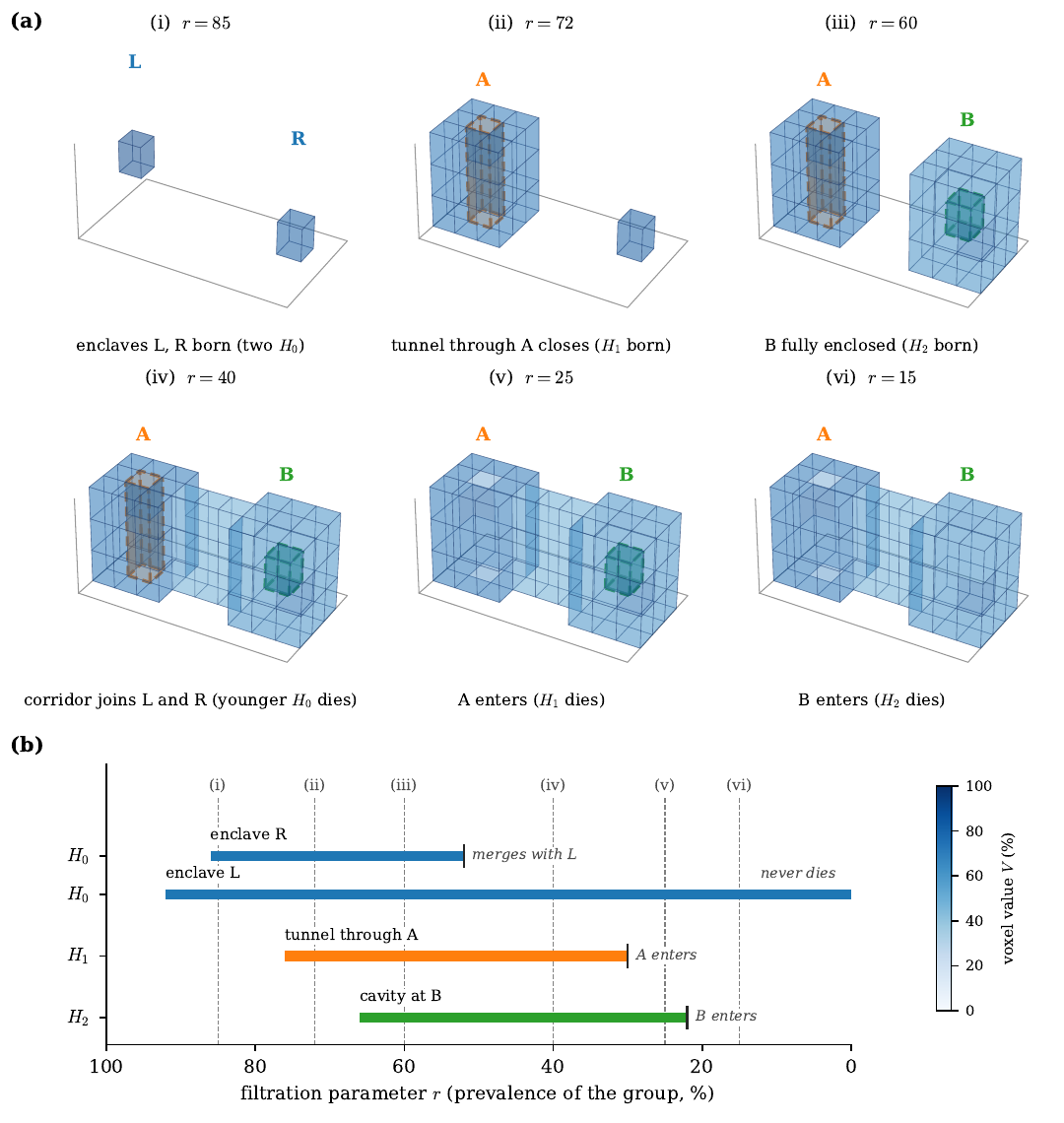}
    \caption{Toy example of a cubical complex filtration where the 0, 1 and 2-persistent homologies are computed and shown in its corresponding barcode.}
    \medskip
    \parbox{\linewidth}{\footnotesize{\textit{Note:} Cells A and B correspond, respectively, to the $H_1$ and $H_2$ situations described in Section~\ref{sec:meaning}. Dashed outlines mark cells not yet in the complex but surrounded by it.}}
    \label{fig:toy_PH}
\end{figure}

The theory outlined in the previous section for $n$--dimensional tetrahedra can be mimicked for $n$--dimensional cubes. This alternative approach, known as the \emph{cubical complex} setting \cite{zomorodian_carlsson,kaczynski_computational, strombom_cubical}, is particularly well suited for the study and analysis of grayscale images \cite{Bleile_dual,choe_cubical_images}. Figure \ref{fig:gray_to_cubical} illustrates how a cubical complex can be obtained from a grayscale image by means of a $2$--dimensional grid. The resulting $2$--dimensional cubes are known as pixels in analogy to digital photography. These inherit the intensity values from the original image, yielding a natural filtration parameter. Pixels are introduced into the filtration process in order of decreasing intensity: darker regions appear first, followed progressively by lighter ones; see Figure \ref{fig:gray_to_cubical}. The same construction extends to three dimensions, where the cubes are called voxels. Figure~\ref{fig:toy_PH} shows a toy example of such a filtration, together with the connected components ($H_0$), loops ($H_1$) and cavities ($H_2$) that appear and disappear along it, and its barcode. We will come back to Figure~\ref{fig:toy_PH} and the specific example it represents in Section~\ref{sec:meaning}. We refer to the Supplementary Information (SI) for a more detailed discussion about cubical complexes.

\subsection{Volume-type complexes for time-series data}{\label{sec:volume}}

To approach population displacement in a manner that can be generalised to other cities, our methodology relies solely on address change data and the city’s administrative division into neighbourhoods. This data undergoes the following preprocessing.
We begin by dividing all individuals who moved in a given year into four groups according to their origin and destination. For each group, we then construct a three-dimensional cubical complex (a $3$-d body whose $3$--dimensional cubes have an associated value that describes their intensity) and compute its persistent homology. These cubical complexes encapsulate both the city geography and the temporal evolution of each population group.

Recall that our dataset spans $20$ consecutive years, where each year's data may be thought of as an origin-destination matrix whose entries are the number of people moving from a given neighbourhood to another area. That is, if the entry in row $i$ and column $j$ is $N$, this means that $N$ people moved from neighbourhood $i$ to neighbourhood $j$.
We simplify these matrices by replacing all the possible destinations by the following summarised four ones: the same as the origin neighbourhood (`stay'); a different neighbourhood within Madrid (`city'); a different town or city in the Comunidad de Madrid, the region Madrid belongs to (`C. Madrid'); and another region or country (`outside'). Hence, in these new reduced matrices, the entry in row $i$ and column $3$ contains the number of people who moved from neighbourhood $i$ to another city in the region of Madrid. Although each of these groups may be further divided, we opt for a consolidated approach to maintain simplicity and analytical efficiency. Finally, we normalise the resulting matrix so that each row sums to $1$.

We construct a $3$--dimensional cubical complex for each of the groups described above so that they reflect the geography of the city neighbourhoods while simultaneously incorporating a temporal component.
To do so, we leverage the fact that data in every year references the same geography up to minor changes; see Section~\ref{sec:data} and SI.
We start by building a $100 \times 100$ grid covering the city map, dividing the city into squares of the form $[x, x+1] \times [y, y+1]$, and then add a third dimension corresponding to the time. Since we have data for $20$ consecutive years, this yields a $100 \times 100 \times 20$ grid. Notice that this grid is independent of the group.

We now make the mapping from the grouped origin--destination data to the voxel values explicit.
Let $\mathcal{B}$ be the set of neighbourhoods of Madrid and $\mathcal{D}$ the set of all possible destinations for that specific origin (the origin neighbourhood, other neighbourhoods in the city, another municipality in the Comunidad de Madrid, or any other region or country).
For each origin, this yields the partition $\mathcal{D} = D^{b}_1 \sqcup D^{b}_2 \sqcup D^{b}_3 \sqcup D^{b}_4$, where
$D^{b}_1=\{b\}$ (`stay'), $D^{b}_2=\mathcal{B}\setminus\{b\}$ (`city'), $D^{b}_3$ the municipalities of the Comunidad de Madrid other than Madrid (`C. Madrid'), and $D^{b}_4$ the remaining destinations (`outside').
We write $M^{(t)}_{bd}$ for the number of people who moved from neighbourhood $b \in \mathcal{B}$ to destination $d \in \mathcal{D}$ in year $t \in \{2004,\dots,2023\}$.
Fix now one of the four groups, $k\in \{ 1, \ldots , 4\}$.
For people leaving neighbourhood $b$ in year $t$, the prevalence of group $k$ is

\[
P^{(t)}_k(b) = \left\lfloor 100 \cdot \frac{ M^{(t)}_{bd}}{\sum_{d \in D} M^{(t)}_{bd}} \right\rfloor \in \{0, 1, \dots, 100\},
\]
where $\lfloor \cdot \rfloor$ denotes the integer part.

Specifically, we take $\mathcal{B}$ to be the set of the 131 current neighbourhoods, reflecting the minor boundary adjustments enacted by the city council in 2017; see SI for the details. Next, we divide the bounding box of the municipality into a grid of $100\times 100$ equal rectangular cells, indexed by $(x,y)\in\{0,\dots,99\}^2$, and let $\nu \colon \{0,\dots,99\}^2 \to \mathcal{B} \cup \{\varnothing\}$ assign to each cell the neighbourhood containing its center, and $\varnothing$ if the center lies outside the municipality; see Section~\ref{sec:cubical}
The vertical coordinate encodes time: the voxel layer $[z,z+1]$, $z\in\{0,\dots,19\}$, corresponds to year $t_z = 2004+z$.
The grayscale image of group $k$ is the map $\mathcal{I}_k \colon \{0,\dots,99\}^2 \times \{0,\dots,19\} \to  \{0, 1, \dots, 100\}$,
\[
\mathcal{I}_k(x,y,z) =
\begin{cases}
P^{(t_z)}_{k}\big(\nu(x,y)\big) & \text{if } \nu(x,y) \neq \varnothing,\\
0 & \text{otherwise.}
\end{cases}
\]
This yields four $3$--dimensional images $\mathcal{I}_1,\dots,\mathcal{I}_4$ of the same size, sharing the same grid.

\begin{figure}[tbhp]
\centering
\includegraphics[width=200pt]{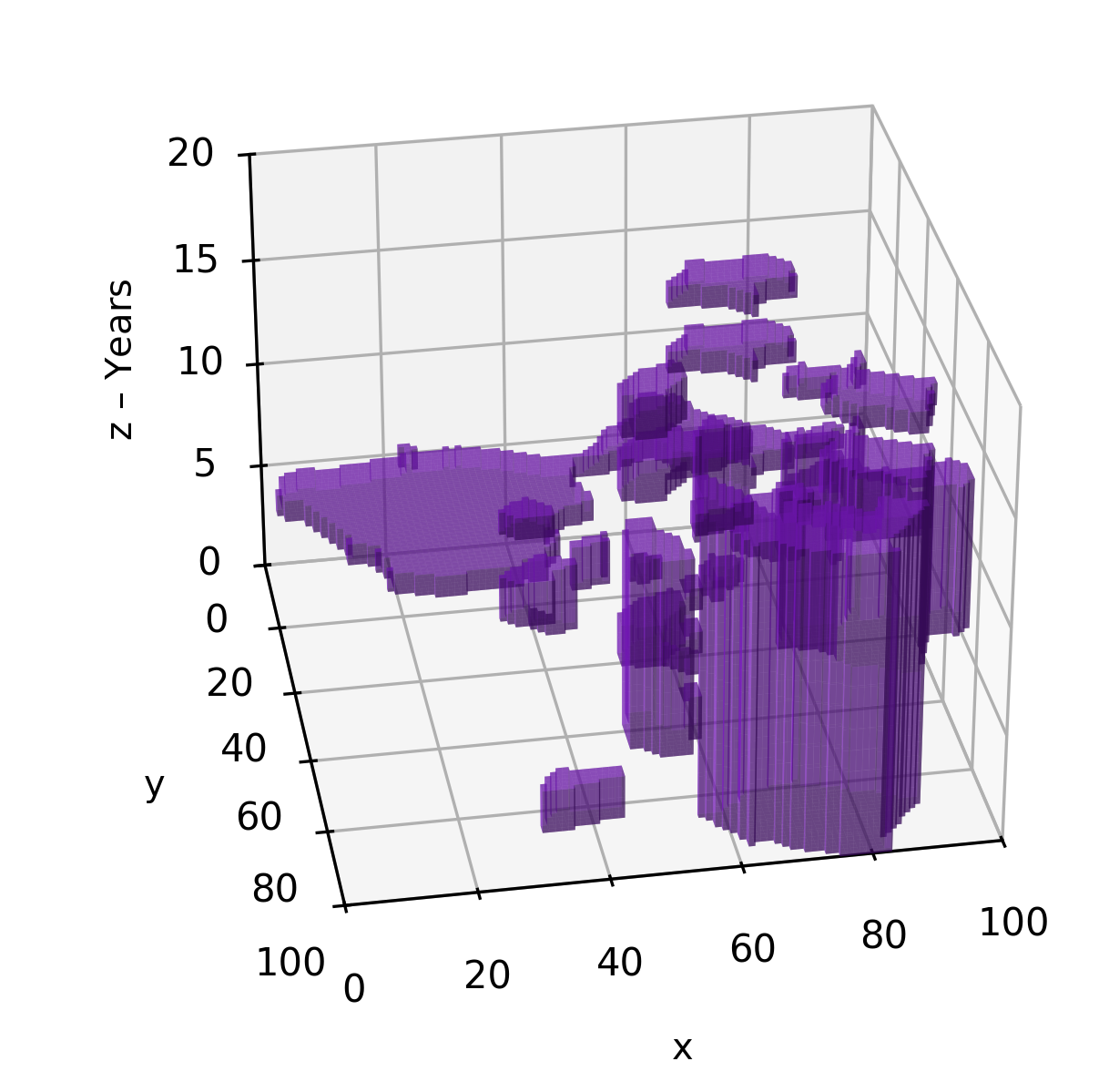}
\caption{Example of the cubical complex for people moving that stayed in the same neighbourhood obtained by selecting the parameter value $r=30$. The city geography prevails on the $xy$-plane, whereas the temporal component is encoded in the vertical axis. The $20$-year period examined corresponds to the range of the vertical axis, with earlier years corresponding to lower values. Derived from neighbourhood boundaries by the Ayuntamiento de Madrid
(\url{https://datos.madrid.es/dataset/300496-0-barrios-madrid}), CC BY 4.0.}
\label{fig:3d_stay}
\end{figure}

We now construct four filtrations. Fix one of the grayscale digital images. 
Its top--dimensional cells or \emph{voxels} are the cubes constructed in the previous paragraph, which correspond precisely to the elements in the $3$--dimensional grid, and their lower-dimensional counterparts are all their faces, edges and vertices.
We then construct a filtration $V$ assigning to voxels the value from the grayscale image, and then extend it top-down to their faces, edges and vertices by assigning to them the smallest value of all the adjacent voxels.
This yields four nested sequences of cubical complexes indexed by a parameter $r$ ranging from $100$ to $0$: recall that for a fixed value $r$, 
a voxel $c$ is selected if $V(c) \geq r$.
The resulting filtered complexes preserve the geography of the neighbourhoods, see Figure \ref{fig:3d_stay}. For a more technical exposition of this construction of the cubical complexes at parameter $r$, known as the T--construction, we refer the reader to the SI. 

\begin{figure}[tbhp]
\centering
\includegraphics[width=350pt]{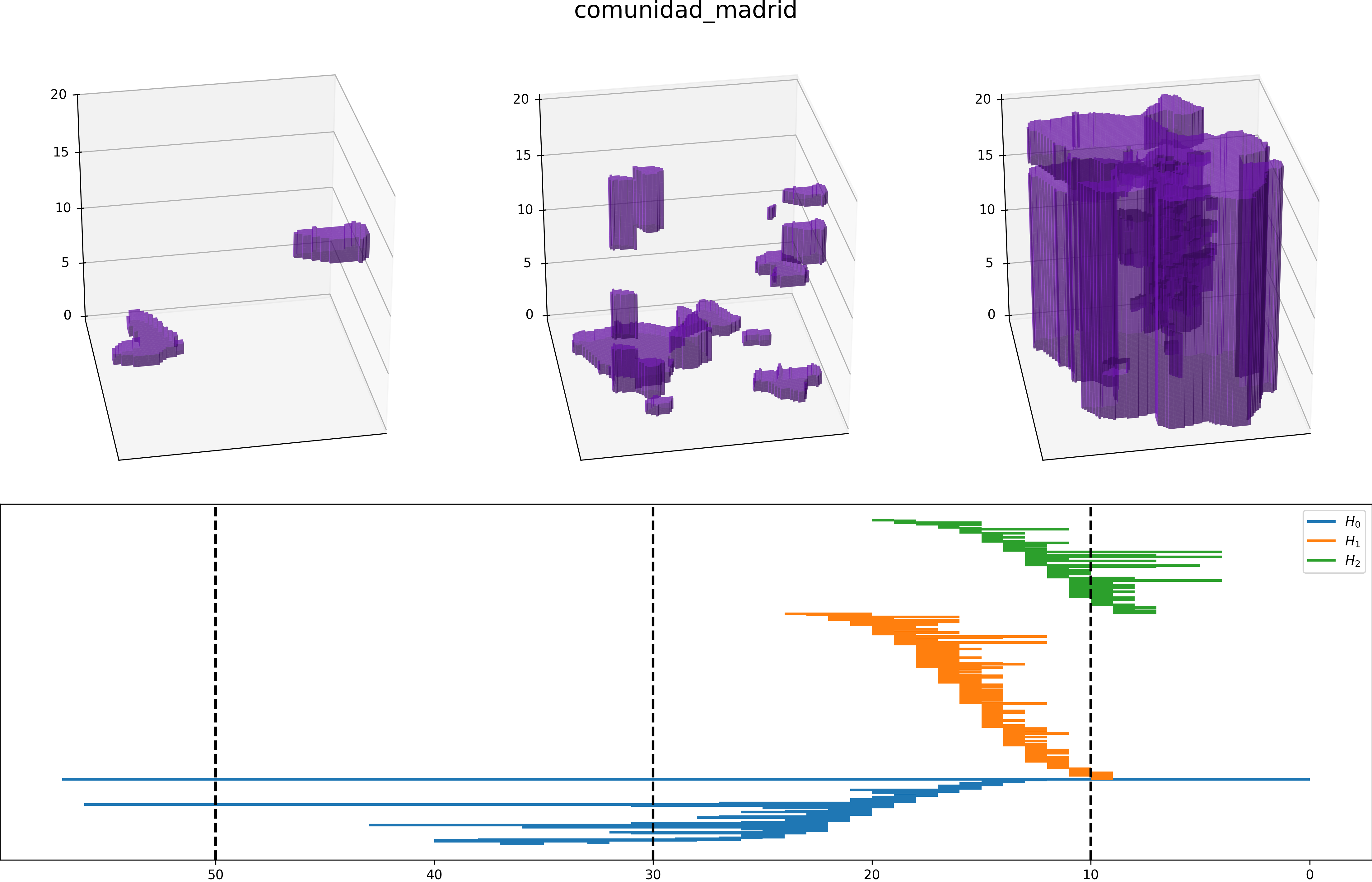}
\caption{Example of developing persistence for the group of people that moved from Madrid to another city within the same region.
The three images at the top row show the cubical complexes for people who moved from Madrid to another city within the same region obtained by selecting the parameter values $r=10$, $30$ and $50$.
The bottom row shows the barcode obtained for this filtered complex, displaying the persistence of the connected components ($H_0$ features),  topological loops ($H_1$ features) and of the cavities ($H_2$ features). Derived from neighbourhood boundaries by the Ayuntamiento de Madrid
(\url{https://datos.madrid.es/dataset/300496-0-barrios-madrid}), CC BY 4.0.}
\label{fig:3d_barcode_comunidad_madrid}
\end{figure}

\begin{figure}[tbhp]
\centering
\includegraphics[width=200pt]{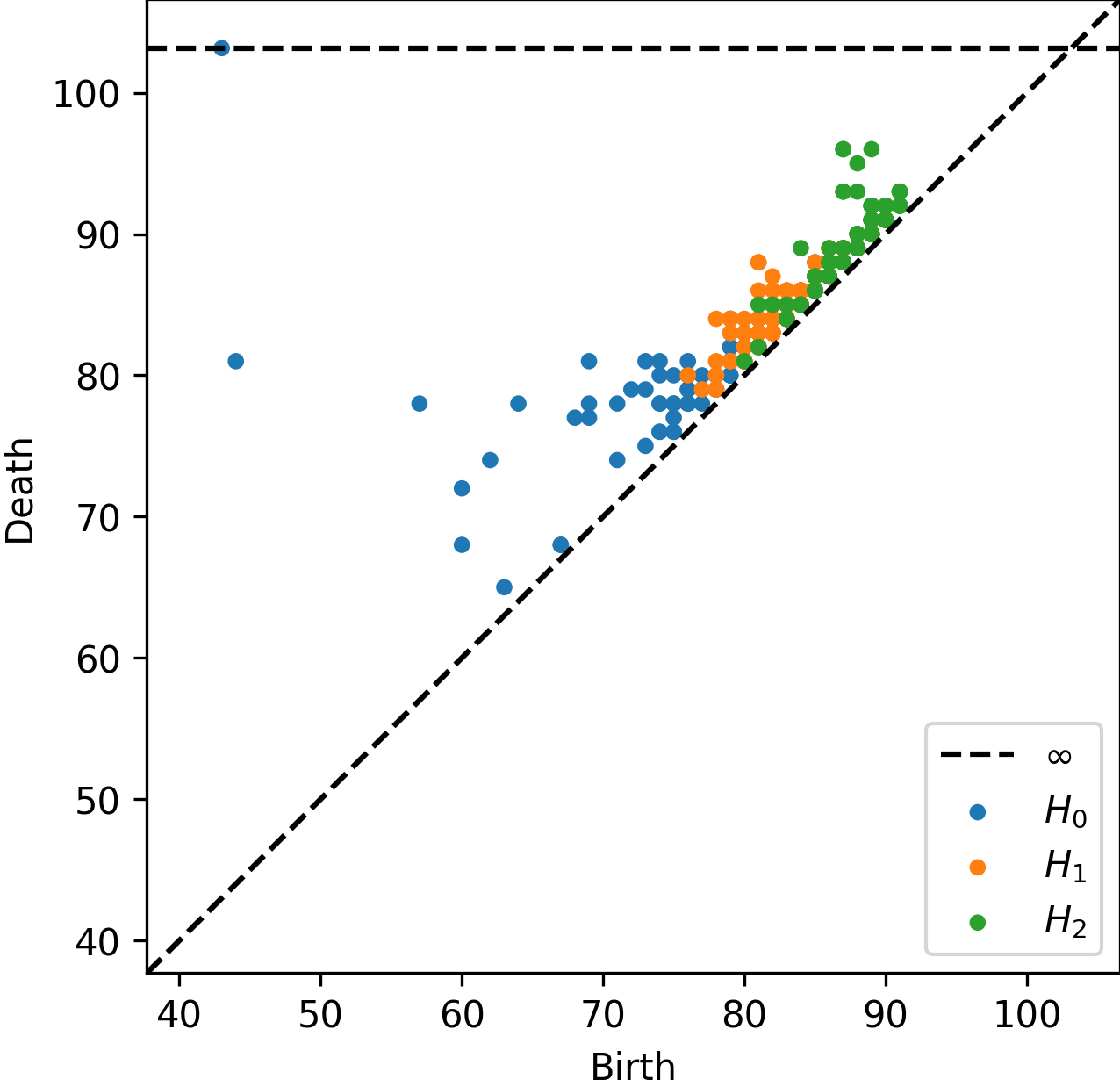}
\caption{Persistence diagram of the filtered complex for the group of people who moved from Madrid to another city within the same region, displaying the $H_0$, $H_1$ and $H_2$ features. Points close to the diagonal present low persistence. Values are shown as $100 - r$, so that birth precedes death.}
\label{fig:pd_comunidad_madrid}
\end{figure}

We then compute persistent homology for each of the four resulting filtrations using CubicalRipser, an extension of the Ripser Python library designed for cubical complexes \cite{cubicalripser}. 
The reason for this choice is three-fold: It supports the construction we chose for cubical complexes, it is robust, and the output format provides not only the birth and death times of the topological features, but also the coordinates $(x,y,z)$ where they initiate and die. We can now represent the obtained persistent homology; see Figure \ref{fig:3d_barcode_comunidad_madrid};  
or its persistence diagram; see Figure \ref{fig:pd_comunidad_madrid}. 

Before turning to the interpretation of the topological features, we illustrate how the abstract construction above, sketched with a toy example in Figure~\ref{fig:toy_PH}, translates to our data. Figure~\ref{fig:3d_stay} shows the cubical complex for the group `stay' at $r = 30$: each purple voxel is a neighbourhood--year in which at least 30\% of the people who moved stayed in their own neighbourhood, whereas empty regions correspond to neighbourhood--years where this share is lower. 

Figure~\ref{fig:3d_barcode_comunidad_madrid} shows how the complex for the group `C.~Madrid' evolves along the filtration. At $r = 50$, only two separate blocks are present, comprising the neighbourhood--years with the highest prevalence of moves to the rest of the region. At $r = 30$, further components have appeared and existing ones have started to grow vertically along the time axis. At $r = 10$, almost all of them have merged into a single body, which nevertheless encloses some empty regions. The barcode records these events. Two $H_0$ bars are much longer than the rest, since these components remain separate over a wide range of $r$. The $H_1$ bars appear only below $r \approx 25$ and are all short. The $H_2$ bars appear below $r \approx 20$, and only a few of them persist noticeably. 

The persistence diagram in Figure~\ref{fig:pd_comunidad_madrid} summarises the same information. Most points lie close to the diagonal, that is, they correspond to short-lived features that can be regarded as noise, and only a few stand out, mainly in $H_0$. This motivates focusing on the most persistent features in each homology degree; see Section~\ref{sec:findings top} and Table~\ref{table:num_features}.

\subsection{Meaning of topological features}{\label{sec:meaning}}

Let us now illustrate how TDA can help us extract such topological patterns. Our approach examines the ``shape'' of the data for each group: `stay', `city', `C. Madrid', and `outside'. In particular, how the residents of Madrid neighbourhoods moved over a 20-year period, by means of the persistent homology of four cubical complexes, one for each group. We therefore explain how the resulting topological features may be interpreted.

Recall that our data is $3$-dimensional, conforming a solid body. This follows from simultaneously considering the geography of the city and the temporal component. The highest-dimensional holes are thus cavities, which may be thought of like the hollow inside a balloon or a soap bubble---an empty region fully enclosed by a surface. In contrast, analysing one year at a time would produce a lower-dimensional complex, and therefore yield a trivial $2$-dimensional homology group $H_2$. Homology groups in dimensions $n \geq 3$ are trivially zero, so we will need to consider all homology groups up to the second. We shall exploit this structural richness in our work.

Consider one of the four groups. Features in $H_0$ correspond to enclaves of people moving to that particular destination. Specifically, if an enclave $\mathrm{A}$ is born at filtration value $r$, it means that the prevalence of that group, e.g. `stay', in that area is $r$. That is, a connected component (a $0$--dimensional hole) born at filtration value $r$ represents an enclave where that group has prevalence $r\%$.
These enclaves may span several consecutive years, possibly comprising different neighbourhoods across that time period.
If a neighbourhood adjacent to an existing enclave $\mathrm{A}$ appears as the parameter $r$ evolves, it is incorporated into $\mathrm{A}$ and no new $H_0$ feature is created. The death time of a feature marks the point at which it merges with an older (i.e., earlier born) enclave. Therefore, persistence captures how distinct and stable an enclave is relative to nearby ones.

Highly persistent $H_0$ features, i.e. those spanning a wide range between its birth and death, are relevant enclaves for that group. In particular, the most prominent $H_0$ feature corresponds to the neighbourhood and year with the highest concentration of that group, which corresponds to its birth time. Since it never merges with another enclave, its death time is $0$, the lowest possible value of the filtration parameter.

While homology classes in $H_0$ correspond to connected components, the ones in $H_1$ correspond to ring-like structures surrounding empty areas in out solid body, and $H_2$ to fully enclosed cavities.
In both cases, these holes represent
regions in space and time where the group is not sufficiently prominent at that step of the filtration.
To illustrate the difference between $H_1$ and $H_2$, consider a neighbourhood $\mathrm{A}$ that is fully surrounded by other city neighbourhoods--not located at the border of the city--together with all its adjacent neighbourhoods, over a period of three consecutive years. On the one hand, if an enclave consists of only all those neighbourhoods except $\mathrm{A}$ over all three years, then we have an $H_1$ feature. We  obtain a ``cylinder''-type construction without lids, i.e., a ring-like structure like a three-floor donut building. On the other hand, if neighbourhood $\mathrm{A}$ is only missing in the middle year, then we have an $H_2$ feature.

The former situation suggests that something in $\mathrm{A}$ is different from its adjacent neighbourhoods, as the hole is maintained over time, and the latter that this phenomenon is restricted to a specific period of time. Analogous examples may be built exchanging the time and space variables.
In general, $1$--dimensional holes point at phenomena that are not bounded to part of the space or to time periods present in the enclave, whereas $2$--dimensional ones indicate that the phenomenon is bounded to a specific area and period. Persistence of $H_1$ and $H_2$ features indicates how long it takes for that hole to be closed. Hence, persistent $H_1$ and $H_2$ features suggest that the neighbourhoods and years belonging to that hole present different characteristics from close-by neighbourhoods and years, which drive individuals to behave differently. 

The two types of holes also differ in the information that can be extracted from them, as illustrated in Figure~\ref{fig:toy_PH}. A cavity ($H_2$ feature) is an empty region fully enclosed by the complex, such as cell B in Figure~\ref{fig:toy_PH}(iii): it is separated from the rest of the empty space, so at every filtration value during its lifetime the set of voxels it encloses, that is, the neighbourhood--year pairs where the group is not yet sufficiently prevalent, is uniquely determined and can be extracted automatically. In contrast, the empty region surrounded by a loop ($H_1$ feature) is open at its ends and therefore connected to the empty space outside the complex, as happens with the tunnel through cell A in Figure~\ref{fig:toy_PH}(ii). In this toy example the tunnel is straight and easily identified by eye, but in general it may bend, e.g.\ running along the time axis through a neighbourhood and then turning horizontally across adjacent neighbourhoods until reaching the city border, or merge with other empty regions; see Figure~\ref{fig:SI_tunnels} in the SI. Hence, there is no well-defined set of voxels associated with an $H_1$ feature. Consequently, for $H_2$ features we can list exactly which neighbourhoods and years are enclosed (see Section~\ref{sec:composition}), whereas for $H_1$ features we report the neighbourhoods and years where they are born and die.

Notice that certain topological values admit an elementary characterization. Indeed, consider a function $f$ on a voxel grid in $\mathbb{R}^3$, assigning to each voxel the prevalence of a given group. In the superlevel-set filtration of $f$, connected components are born at local maxima. However, their death times depend on how components merge as the threshold varies, which requires information about the global connectivity of the superlevel sets. Specifically, a component may merge with younger components before merging with an older one, and only the latter merger determines its death time, under the elder rule, and hence its persistence.
For $H_2$, the death times of finite features correspond to local minima of $f$, since a cavity disappears when its last remaining voxel is added, 
but not every local minimum necessarily corresponds to an $H_2$ death, and their birth times cannot generally be identified from local extrema. 
In $H_1$, neither birth nor death times admit comparable local characterizations. Thus, although certain birth and death times can be identified through elementary properties of $f$, the persistence of topological features cannot, in general, be recovered from these properties alone.

\begin{figure}[!h]
\includegraphics[width=0.8\textwidth]{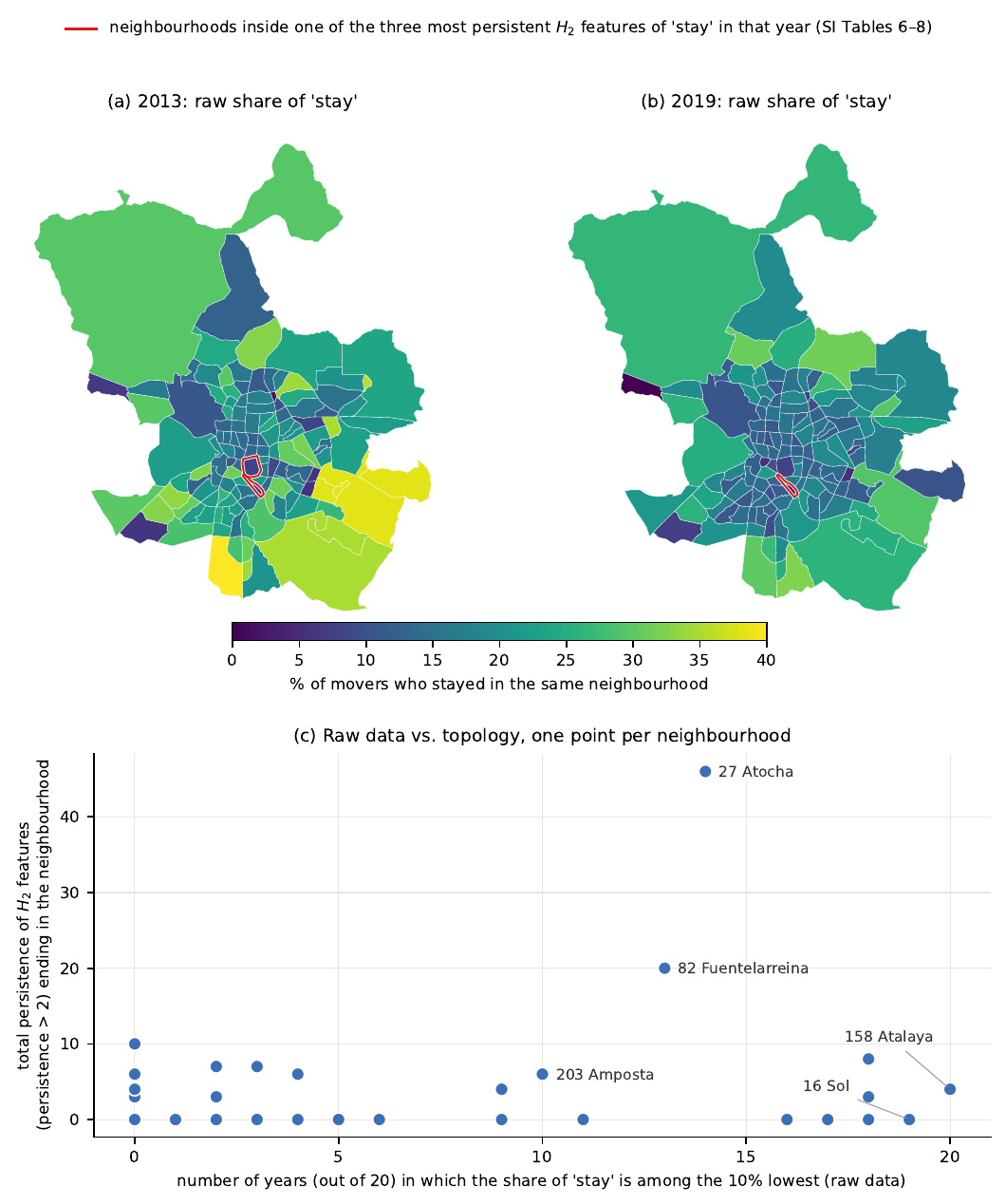} 
\caption{Raw data and topological features for the group `stay'. Derived from neighbourhood boundaries by the Ayuntamiento de Madrid
(\url{https://datos.madrid.es/dataset/300496-0-barrios-madrid}), CC BY 4.0.}
\medskip
\parbox{\textwidth}{\footnotesize\textit{Note:} Panels (a) and (b) show, outlined in red, the neighbourhoods enclosed in 2013 and 2019 by the three most persistent $H_2$ features of the group `stay' (SI Tables~6--8). In panel (c), each point is a neighbourhood. The horizontal axis gives the number of years, out of 20, in which its share of stay' is among the 10\% lowest of that year. The vertical axis gives the sum of the persistences of the $H_2$ features with persistence greater than 2 that die in that neighbourhood, that is, whose cavity is closed when that neighbourhood enters the filtration; each feature is counted for a single neighbourhood. Many neighbourhoods share the same values, so their points overlap; most of them lie at $(0, 0)$.}
\label{fig:raw_vs_topology}
\end{figure}

Finally, a natural question remains: what does this method, through its encoding in terms of topological features, reveal that cannot be obtained directly from the raw data? Figure~\ref{fig:raw_vs_topology} compares these results with the raw data. Panels~(a) and~(b) show the share of the `stay' group in 2013 and 2019 respectively.  The neighbourhoods enclosed by the three most persistent $H_2$ features of `stay' in those years are highlighted in red (SI Tables~6--8): `27 Atocha' and `35 Los Jerónimos' in 2013, and `27 Atocha' in 2019. Although the prevalence of the group is low there, the maps show that so is it in many other neighbourhoods. In fact, in any given year, the 10\% lowest shares are spread across the city, and several neighbourhoods remain among them in almost every year of the period (e.g. `158 Atalaya' in all 20 years and `16 Sol' in 19). A yearly map or ranking therefore does not single out any particular area. Panel~(c) makes this comparison quantitative. For each neighbourhood, it plots the number of years in which its share is among the 10\% lowest against the total persistence of the $H_2$ features (persistence $> 2$) that end in it, that is, whose last neighbourhood-year to enter the filtration lies in that neighbourhood. Neighbourhoods that are persistently low in the raw data obtain little or no persistence, whereas `27 Atocha', among the 10\% lowest in 14 years, concentrates by far the largest total persistence (46, compared with 20 for the next neighbourhood, 82 Fuentelarreina'). Atocha is also the neighbourhood-year that fills, and thus closes, the three most persistent $H_2$ features of stay'. Persistence thus does not measure how low a share is, but how much lower it is than its surroundings in both space and time, and this is what distinguishes Atocha from the rest.

Since the methodology presented in this article may be challenging for readers less familiar with Topological Data Analysis or persistent homology, a flowchart summarising the full pipeline is provided in Figure~\ref{fig:pipeline} of the Supplementary Information.

\subsection{Post-processing of the topological features}\label{sec:composition}

We examine the topological features obtained for each of the four groups. 
TCripser provides the spatial–temporal coordinates xyz of the birth and death of each feature, allowing us to map them back to specific neighbourhoods and years.  
While birth and death places are directly informative for $H_0$ features, further developments are needed to fully exploit the topology of $3$--dimensional grayscale images.
Specifically, we extract the neighbourhoods and years that constitute each connected component \emph{at a given filtration-parameter value}.

To gain insight into displacement patterns from $2$--dimensional features, we identify which neighbourhoods and years are enclosed within each cavity at its birth.
TCripser outputs require a cleaning process, as duplicate entries frequently appear in $H_{1}$ and $H_{2}$. Additionally, our development revealed that some reportedly ``distinct" cavities--associated with different birth or death voxels--enclosed indeed identical sets of voxels.
All such duplicates have been removed from the analysis.

\subsection{Description of the topological features}{\label{sec:findings top}}

For each of the four groups, we construct a suitable space and analyse its most persistent topological features, since they provide the strongest signals. We focus on the three most persistent ones of each homology degree. 
Prevalences are truncated to integer percentages, thus all birth and death values, and hence all persistences, are integers. 
Persistences of 1 may arise solely from this truncation procedure.
We therefore regard as important those features with persistence greater than 2, which guarantees a difference of more than two percentage points in the underlying prevalences, i.e., more than twice the resolution of the data; Table~\ref{table:num_features} reports their number for each group and homological degree.

The SI reports birth and death neighbourhoods, years, and prevalence at birth time (share values) for the ten most prevalent topological features in each group and homological degree (0–2). Additionally, tables 6-17 in SI list the neighbourhood-years enclosed in the three most persistent cavities at birth time in each group, together with the prevalence of that group in each neighbourhood-year.

Let us first analyse the $H_0$ features, which represent neighbourhood–year combinations where a given group is particularly prominent.
In the group `stay', the most persistent feature originates in 2009 in neighbourhood `27 Atocha', located between the main train station Atocha and the M\'endez \'Alvaro transportation hub.
While this feature requires a while to expand, others born around the same time quickly span several years and neighbourhoods. 
This highlights a difference in the population dynamics compared to the surrounding areas and years. 
Nevertheless, only at high values of the filtration-parameter do the connected components in this group expand vertically beyond the 2020 threshold or include neighbourhoods in the `01 Centro district', the historic centre. This suggests that residents in this district face difficulties relocating within their own neighbourhood, a pattern that appears to extend to the wider city from 2020 onward.

\begin{figure}[tbhp]
\centering
\includegraphics[width=200pt]{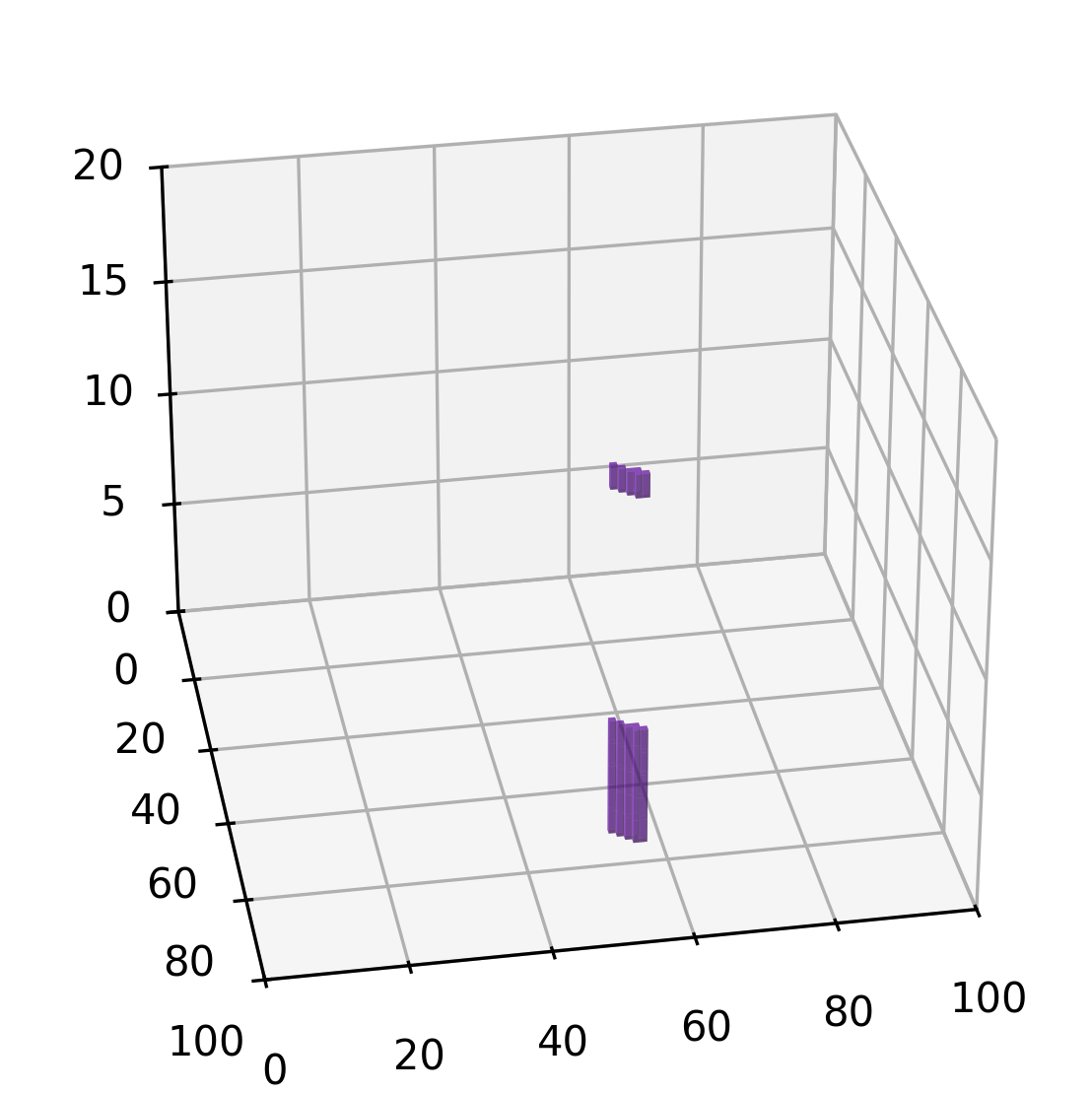}
\caption{First two connected components born in the group `city'. Derived from neighbourhood boundaries by the Ayuntamiento de Madrid
(\url{https://datos.madrid.es/dataset/300496-0-barrios-madrid}), CC BY 4.0.}
\medskip
\parbox{330pt}{\footnotesize\textit{Note:} Both components cover only the neighbourhood 27 Atocha and span two non-consecutive sets of years. This can be observed directly from the graph: the $x$ and $y$ coordinates remain unchanged, while the $z$ coordinate varies and some values are not attained.}
\label{fig:city_threshold_70_}
\end{figure}

The `city' group exhibits a larger number of connected components than the others (see Table \ref{table:num_features}), suggesting that this group is more dispersed across space and time. Its two most persistent $H_0$ features both arise in neighbourhood `27 Atocha' and expand only later, forming stacked components interrupted in 2009–2010--coinciding with the emerging of the `stay' $H_0$; see Figure \ref{fig:city_threshold_70_}.
This pattern suggests that, when conditions allow--such as in the period following the 2008 financial crisis--individuals tend to prefer moving within their own neighbourhood rather than relocating elsewhere in the city.

\begin{table}[h!]
\centering
\begin{threeparttable}
\begin{tabular}{llrrrr}
 &  & stay & city & C. Madrid & outside \\
\midrule
$\#$ features& $H_0$&73&98&80&75\\
&$H_1$&209&211&197&201\\
&$H_2$&120&79&105&132\\
\midrule
$\#$ important features &$H_0$&35&54&34&39\\
&$H_1$&66&84&23&44\\
&$H_2$&38&24&13&31\\
\bottomrule
\end{tabular}
\par\smallskip
{\footnotesize\noindent\textit{Note:} Recall that a $0$-dimensional feature ($H_0$) is a connected component, a $1$-dimensional feature ($H_1$) is a loop, and a $2$-dimensional feature ($H_2$) is an enclosed cavity. Hence, the number of connected components in each group is given by the $H_0$ rows.\par}
\caption{Number of topological features in each group. Important features are those with persistence $>2$.}\label{table:num_features}
\end{threeparttable}
\end{table}

In the `Comunidad de Madrid' group, the first connected components are born at the city border, in neighbourhoods that are spatially distant from one another.
They initially expand to further years, i.e., along the vertical axis, and thus stay distant, resulting in high persistence as they do not merge. Given their locations, these patterns do not necessarily indicate significant displacement, as residents may simply relocate within the immediate vicinity, but may instead reflect short-distance relocations within nearby areas.
In contrast, $H_0$ features in the `outside' group are more indicative of displacement. Neighbourhood `141 Pavones (East)' appears to experience high pressure, as several of the most persistent features are born there. 
Furthermore, the rapid, citywide expansion of highly persistent features after 2016 likewise suggests that displacement has intensified and become more widespread in recent years.

Higher dimensional topological features contribute 
to the analysis of patterns potentially indicative of displacement as follows: $1$-- and $2$--holes in the groups `stay' and `city' suggest displacement, whereas those in `Comunidad de Madrid' and `outside' a stable market, especially if accompanied by a `stay' or `city' $0$--feature contained or largely overlapping the cavity. As explained in Section~\ref{sec:meaning} (see Figure~\ref{fig:toy_PH}), $H_2$ features convey more comprehensive information on displacement than $H_1$ ones: since the enclosed neighbourhood--year pairs may be explicitly extracted, we can identify precisely which areas and periods are potentially affected.

All four groups have a high number of low persistent $H_1$ features. The group `stay' presents the most persistent $1$--hole,
whose birth and death hint at population displacement in the north-eastern part of the city (districts `20 San Blas-Canillejas' and `15 Ciudad Lineal') in 2016-2019. The `city' group's two most persistent $H_1$ features both emerge and dissolve near the airport (district `21 Barajas'), revealing patterns compatible with displacement in overlooked peripheral areas.
The $H_1$ features in the `Comunidad de Madrid' group are spatially and temporally spread-out, unlike the `outside' group ones.
The latter, specifically the feature located in the highly touristified `01 Centro' district during the 2020-2021 COVID pandemic,
exemplifies a housing market relaxation and demonstrates our approach's ability to capture subtle population displacement-compatible patterns.

We now examine $H_2$ features, which may exhibit complex dynamics. Note that cavities often emerge as ``nested bubbles" or arise when a single cavity splits into two. Although intuitively similar, these features have different birth and death times. 
To determine whether two cavities are nested, we extract the neighbourhood-years comprising each cavity at birth (see Section~\ref{sec:composition}) and check whether one set is contained in the other.

\begin{figure}[ht]
\centering
\includegraphics[width=300pt]{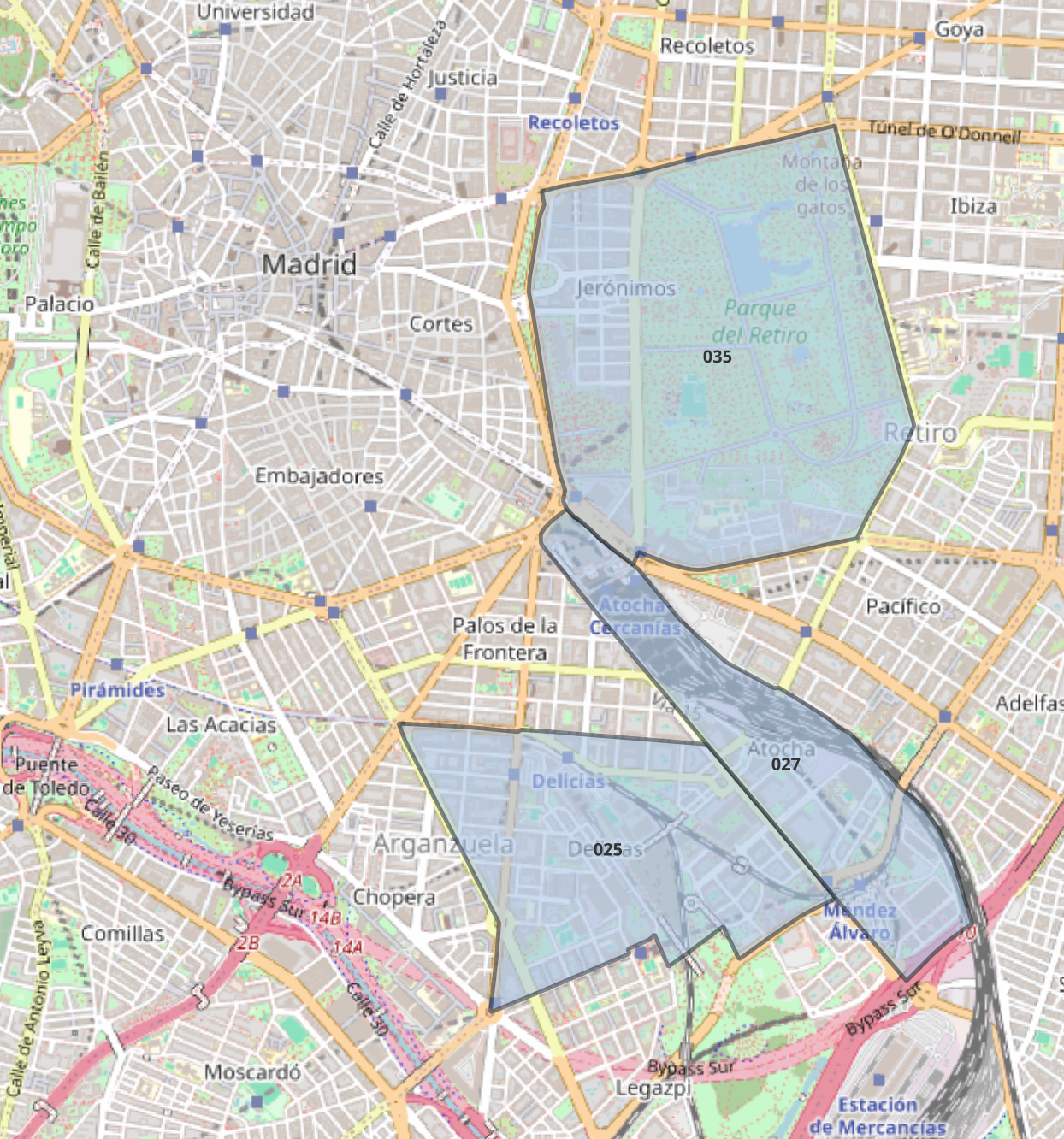}
\caption{Map of central-south Madrid with neighbourhoods 25, 27 and 35 highlighted. Base layer: \href{https://www.openstreetmap.org/}{Open Street map}, licensed under the \href{https://opendatacommons.org/licenses/odbl/}{Open Data Commons Open Database License}. Top layer: Ayuntamiento de Madrid, ``Barrios municipales de Madrid''
(\url{https://datos.madrid.es/dataset/300496-0-barrios-madrid}), CC BY 4.0
(\url{https://creativecommons.org/licenses/by/4.0/}). Map created by the authors.}
\label{fig:selected_neighbourhoods}
\end{figure}

The three most persistent $H_2$ features in the group `stay' intensely affect neighbourhood 27, and also neighbourhoods 25 and 35, from 2013 onwards; see Figure \ref{fig:selected_neighbourhoods}.
The second and third most persistent features both involve the same years in neighbourhood 27, but differ in their spatial extent: one also includes neighbourhood 25, while the other includes neighbourhood 35.
We conclude that the results provide strong evidence suggestive of population displacement in the areas around the Atocha train station since 2013, coinciding with the post-2008 crisis recovery.

The group `city' presents the most persistent $2$--hole of all groups. It encloses one single neighbourhood year pair, namely neighbourhood 27 Atocha in 2018.
The second most persistent feature is particularly notable, as it is confined to neighbourhood 27 during 2020-2021, signaling a bounded decrease in housing market tension in a highly affected area, likely triggered by the pandemic. This illustrates the method's ability to identify both large-scale and localised, temporary population dynamics. The third most persistent one is located in the South-East, in the Moratalaz district, and encloses only neighbouhood 141 Pavones in year 2020, witnessing the effect of the pandemic in suburban areas. 

The $H_2$ features in the `Comunidad de Madrid' and `outside' groups suggest areas and periods of reduced displacement, and both affect predominantly neighbourhood 27 `Atocha'. However, contrary to the other two groups, they concentrate in the first part of the time series.
Specifically, in the `outside' group these three features enclose only neighbourhood `27 Atocha', in years between 2007 and 2019, offering additional evidence of the unique dynamics characterising this area.
The first and third most persistent features in the `Comunidad de Madrid' group are very similar, whereas the second encompasses only neighbourhood 158 `Atalaya' in 2012, in the North-East part of the city; see Figure \ref{fig:mapa_no_displacement}.

The analysis of the topological features of the four groups revealed particularities of patterns suggestive of displacement which may prove more challenging to extract directly from the  raw data.
We found that neighbourhood `27 Atocha' is a key case study, and presents a different behaviour from its surroundings throughout the 20-year period,
demonstrated by persistent $H_0$ and $H_2$ features and cavities consisting only of this neighbourhood. 
Initially part of a stable market, marked by persistent $H_2$ features in `Comunidad de Madrid' and `outside' and a following connected component in `stay' arising after the financial crisis,
its population dynamics shifted dramatically afterwards.
General behaviour compatible with displacement, marked bypersistent cavities in `stay' spanning years  2013-2019 covering it, was briefly interrupted during the COVID-19 pandemic, as shown by presistent `city' cavities. This highlights the neighbourhood's distinct responsiveness to market forces, 
which distinguishes it from its surroundings and acts as a focal point for broader area dynamics.

\begin{figure}[ht]
\centering
\includegraphics[width=\textwidth]{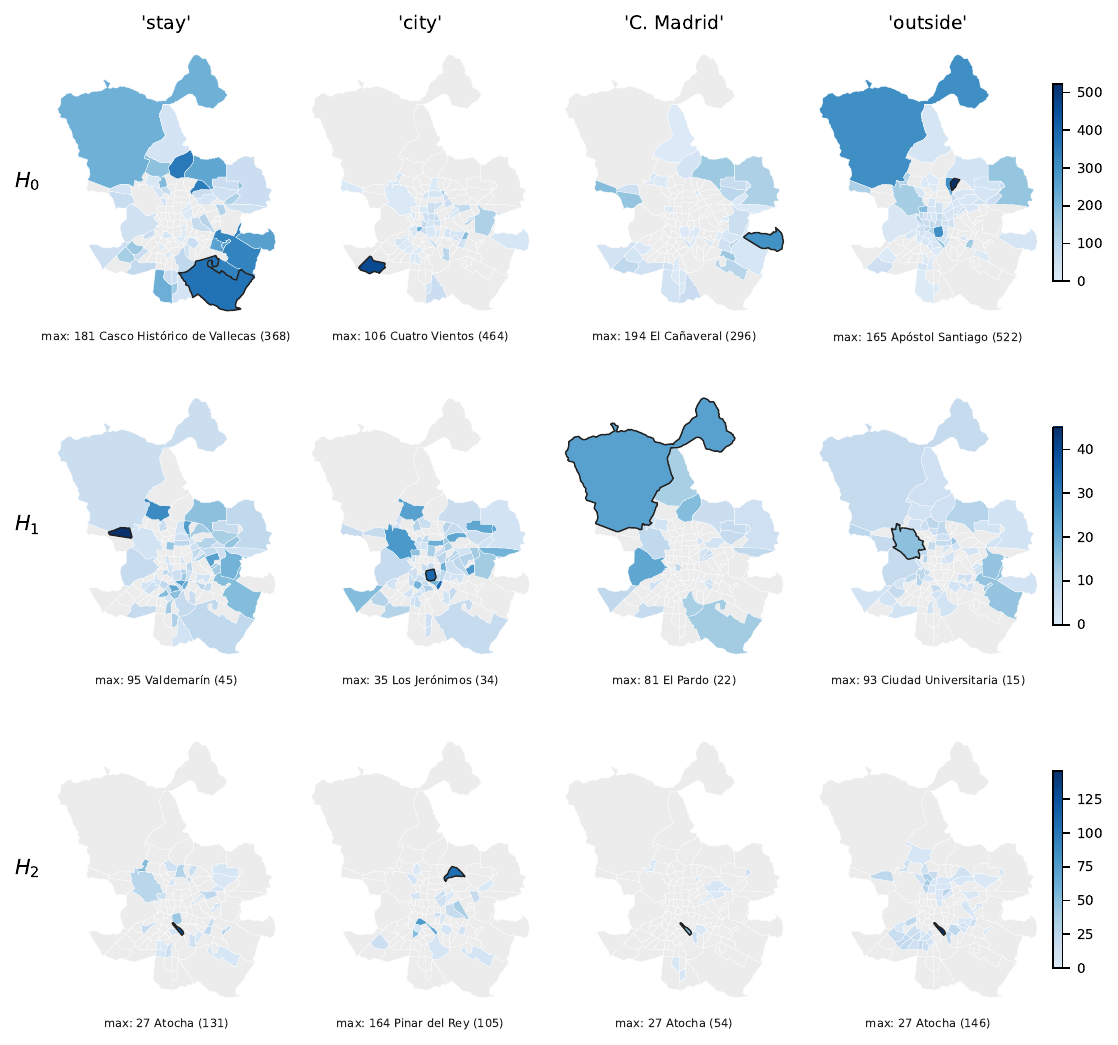}
\caption{Neighbourhood-level summary of the topological features. For each group (columns) and homological degree (rows), sum over 2004--2023 of the persistences of the features with persistence greater than 2 whose region contains the neighbourhood ($H_0$: connected component; $H_1$: volume-optimal cycle; $H_2$: cavity). Grey: 0. The neighbourhood with the largest value in each panel is outlined and named below it. Derived from neighbourhood boundaries by the Ayuntamiento de Madrid (\url{https://datos.madrid.es/dataset/300496-0-barrios-madrid}), CC BY 4.0.}
\label{fig:statistic_map}
\end{figure}

To summarise these results visually, we define a neighbourhood-level statistic, the accumulated persistence, as the sum of the persistences of the features of a given group and homological degree with persistence greater than 2 associated with a neighbourhood over 2004--2023, that is, the connected components ($H_0$) and cavities ($H_2$) containing it, and the loops ($H_1$) born or dying in it. Figure~\ref{fig:statistic_map} maps this statistic for each group and homological degree, thus highlighting the areas where the behaviour of each group departs most from that of neighbouring areas and years.

\section{Results}

\subsection{Identifying potentially displaced individuals}

Displacement is typically defined in terms of residents’ reasons for moving. This information is difficult to obtain directly and is therefore often inferred from socio-economic indicators.
Rather than relying on such aggregated measures, which can obscure important heterogeneity, we adopt a topological perspective.
To this end, we first divide all individuals who moved in a given year into four groups.

Groups are determined by the origin and destination of each move. The underlying assumption is that individuals develop social and practical ties in their area of residence, such as places of study, family and friends, and distance tends to weaken that social network \cite{friendship,residentialmob}. Consequently, individuals who relocate are more likely to prefer to remain near their previous area of residence.
We operationalise this preference by assuming that the most likely option is to move within the same neighbourhood (`stay' group, see Section~\ref{sec:volume}), followed by relocation to another neighbourhood within the same city (`city').

We now illustrate the case of individuals leaving the city. The housing market in Madrid is under significant pressure: according to official statistics, between December 2019 and December 2024 the average rent price per square metre in Madrid has increased $24,64\%$, see \cite{bancodedatosmadrid} section `E. Edificación y Vivienda'.
It is well-established that housing prices impact migration flows. For instance, a study on migration and housing markets conducted by the OECD (Organization for Economic Co-operation and Development)
concluded that high housing costs discourage both remaining and moving into a region \cite{oecd}.
This evidence suggests that the search for more affordable housing is a key driver of out-migration, allowing us to infer population displacement.
We distinguish two possible destinations for those leaving the city: relocating to another town within the same region (`C. Madrid'), or to another region (`outside'). The former is assumed to be more likely, as it enables individuals to maintain some of their social and economic ties, such as employment.
Nevertheless, notice that we do not imply that all individuals moving outside the city do so involuntarily, but rather that an increase in such moves, or a departure from trends observed in neighbouring areas and years, constitutes a pattern consistent with population displacement.

\subsection{Findings}{\label{sec:findings}}

For each of the four groups, we construct a suitable space and analyse its topological features. 
This analysis revealed broader spatial and temporal patterns consistent with population displacement that are not apparent from yearly maps or rankings of the raw data (Figure~\ref{fig:raw_vs_topology}). Importantly, we do not claim to identify displaced individuals, but rather to highlight broader spatial and temporal dynamics at the city level.
This analysis revealed broader spatial and temporal patterns consistent with population displacement that are not apparent from yearly maps or rankings of the raw data (Figure~\ref{fig:raw_vs_topology}). Importantly, we do not claim to identify displaced individuals, but rather to highlight broader spatial and temporal dynamics at the city level.

Our analysis reveals distinctive spatial and temporal dynamics consistent with population displacement in neighbourhood `27 Atocha' (SI Tables 2–5), located between the main train station Atocha and the M\'endez \'Alvaro transportation hub.
Over the 20-year period, it consistently behaves differently from surrounding areas, repeatedly appearing as a focal point across all four groups.
In the earlier years, it reflects relatively stable conditions, characterised by a low presence of moves to the wider region (Comunidad de Madrid) and, from 2009 onward, by a predominance of local movements within the neighbourhood and the city (third most persistent $H_2$ feature of 'C. Madrid' and most persistent $H_0$ feature of 'stay'; SI Tables 4 and 2). However, its dynamics shift markedly thereafter.  From 2013 to 2023, strong patterns emerge suggesting increasing difficulties for residents to relocate within the same neighbourhood, consistent with population displacement.
At the same time, this neighbourhood exhibits short-lived periods of reduced pressure, notably during the COVID-19 pandemic, highlighting  the neighbourhood's distinct responsiveness to market forces, 
which distinguishes it from its surroundings and acts as a focal point for broader area dynamics.

The findings also suggest that from 2013 onward, it becomes increasingly difficult for residents in central districts  to relocate within their own neighbourhoods.
Instead, we observe a growing tendency to move out of the region from these areas.
These patterns provide strong indications of population displacement in central Madrid since 2013,
coinciding with the post-2008 crisis recovery, with the areas around Atocha being most impacted.
Notably, a temporary easing is observed in central areas, including the highly touristified `01 Centro' district, during 2020–2021, likely associated with the COVID-19 pandemic.

We also observe patterns consistent with population displacement in more peripheral areas.
In the north-east of the city--specifically in the districts of San Blas-Canillejas (20) and Ciudad Lineal (15)--residents experienced similar difficulties in moving within their own neighbourhoods during 2016–2019 (most persistent $H_1$ feature of 'stay'; SI Table 2). In the airport area (Barajas, district 21), constraints are observed even for moves within the city (two most persistent $H_1$ features of 'city'; SI Table 3).

Neighbourhood `141 Pavones' stands out as an area under pressure, characterised by a high prevalence of moves leaving the region (SI Table 5).

\begin{figure}[tbhp]
\centering
\includegraphics[width=300pt]{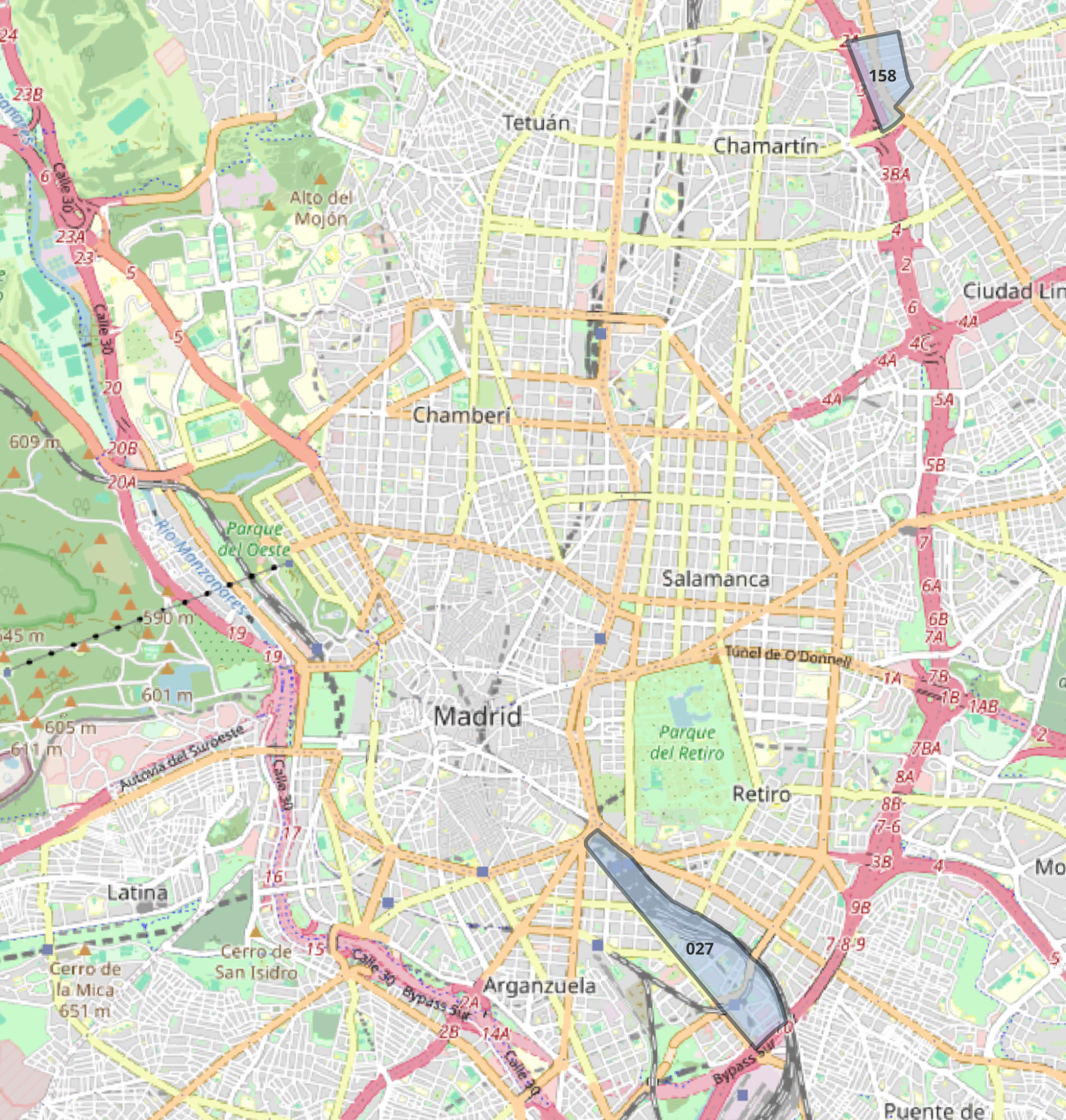}
\caption{Map of Madrid with neighbourhoods 27 and 158 highlighted. Base layer:  \href{https://www.openstreetmap.org/}{Open Street map}, licensed under the \href{https://opendatacommons.org/licenses/odbl/}{Open Data Commons Open Database License}. Top layer: Ayuntamiento de Madrid, ``Barrios municipales de Madrid''
(\url{https://datos.madrid.es/dataset/300496-0-barrios-madrid}), CC BY 4.0
(\url{https://creativecommons.org/licenses/by/4.0/}). Map created by the authors.}
\label{fig:mapa_no_displacement}
\end{figure}

By contrast, the neighbourhood `158' Atalaya, located in the north-east of the city shows periods of relatively weaker indications of population displacement, indicated by bounded intervals in which moves to other towns in the region are less frequent (three most persistent $H_2$ features of 'outside'); see Figure \ref{fig:mapa_no_displacement}.

\section{Discussion}

This study presents a novel quantitative approach to investigating population displacement, a major toll in social urban transformations,
and demonstrates through a case study of the city of Madrid, Spain, that its findings are consistent with documented patterns of displacement.
In particular, it reveals patterns in residential destination choices that align with previously identified areas and periods of displacement.

This method involves two main phases: the construction of suitable cubical complexes and the subsequent analysis of that complex through persistent homology, a widely-used tool from Topological Data Analysis to synthesise the shape of data, particularly robust to noise. 
We leverage publicly available data--the geography of the administrative division of the city into neighbourhoods and address change data (yearly origin and destination of the city inhabitants moving)--to build the grayscale images underlying our cubical complexes.

Prior attempts to quantify population displacement present both data and methodological limitations. Studies relying on ad-hoc metrics, surveys, or tailored to the city specifics hinder comparability, and oftentimes data availability restricts the analysis to short time spans. In contrast, our study only uses publicly available, periodically collected data, enabling replicability across cities, which, combined with persistent homology of the four cubical complexes, allows us to work on an adequate time scale.  Nevertheless, the high temporal and geographical coverage of this data, which encompass the entire population rather than a sample, comes at the expense of the richer information provided by surveys, which often reveal the particular reasons behind individual moves and thus allow specific instances of displacement to be identified.

Methodologically, displacement studies often depend on a subjectively chosen control group, which strongly impacts their conclusions, and rely on aggregate statistical indicators that reduce a complex, multidimensional phenomenon to a single summary value. Our approach overcomes these shortcomings by incorporating all neighbourhoods and all years into a single cubical complex, allowing persistent homology to extract topological patterns--richer than single numerical indicators--based on spatio-temporal proximity without requiring a control group.

Although our approach offers several advantages, it also presents some limitations. First, due to the varied data formats and structures used by national public administrations, data preprocessing may vary by country, consequently impacting the time required to construct the corresponding cubical complexes. Second, the methodology does not, by itself, yield interpretative conclusions: as detailed in Section~\ref{sec:findings top}, once persistent homology is computed, a careful analysis of the resulting features is necessary to relate them to the underlying urban dynamics. Finally, our approach lacks a systematic methodology to reconcile persistent diagrams of different groups in order to extract conclusions.

Further variations of this method may be considered. Looking forward, it would be interesting to split the `city' group into those who move to an adjacent neighbourhood and to the rest of the city. A similar refinement could be applied to the `outside' group by distinguishing moves to adjacent regions, or even other countries.
It would also be interesting to explore this approach at a finer temporal resolution to determine whether additional patterns emerge, although excessively high frequencies (e.g., monthly) may introduce topological features driven primarily by seasonality.
Furthermore, it would also be interesting to study the socioeconomic features of the identified displaced population to enrich the persistent homology study. 
Finally, it would be beneficial to repeat the analysis in another city to further validate this approach.

\section{Acknowledgements}

The authors would like to thank the Department of Mathematics and Quantitative Methods at CUNEF Universidad, Madrid, Spain, for its support throughout the development of this paper.

\printbibliography

\appendix

\newpage

\begin{center}
    \Large Supporting Information
\end{center}

\setcounter{figure}{0}
\setcounter{table}{0}
\renewcommand{\thefigure}{S\arabic{figure}}
\renewcommand{\thetable}{S\arabic{table}}

\newpage

\section{Data}

\subsection{Datasets} 

The main data source for this study is the official demographic data of the city of Madrid. It is publicly available at \cite{bancodedatosmadrid}, in CSV format. We used several tables, specifically those containing information about address changes. It can be found in the `C.Demograf\'ia y Poblaci\'on' section,
under `Padr\'on municipal/Din\'amica geogr\'afica/Cambios de domicilio en la ciudad'. 
We want to work with the finest spatial granularity possible.
However, it is a requirement for us to know where individuals move from and to, as opposed to just the number of individuals arriving/leaving an area. We found that the latter is available at census tract level, whereas the former only at neighbourhood level.
Therefore, we choose the tables in the aforementioned path containing data at neighbourhood level, and consider the whole period of time for which they are available: from 2004 until 2023.

\begin{figure}
\centering
\includegraphics[width=400pt]{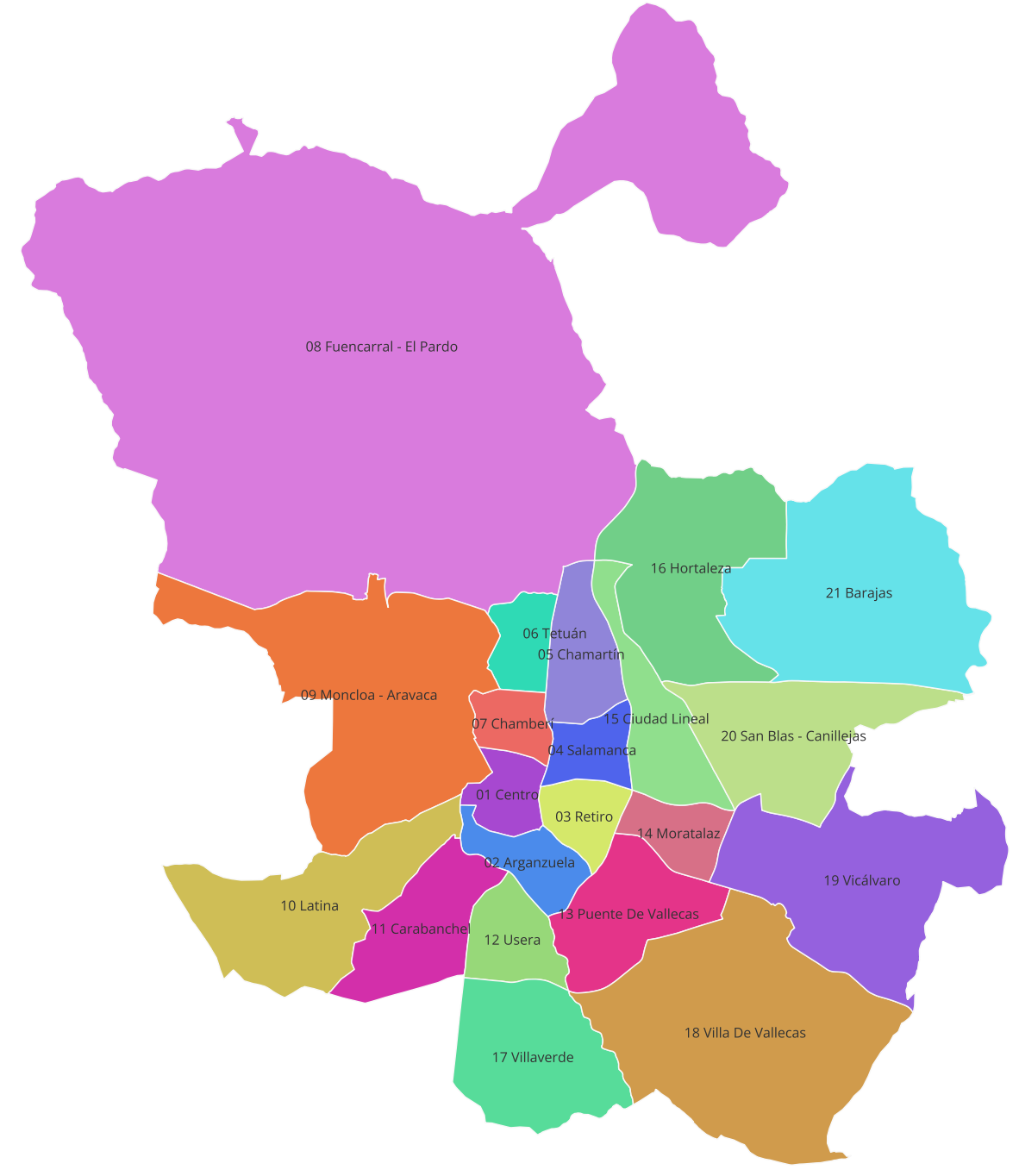}
\caption{Madrid map and their 21 districts as of April 2025. \textcolor{olive}{Base layer: Ayuntamiento de Madrid, ``Distritos municipales de Madrid''
(\url{https://datos.madrid.es/dataset/300497-0-distritos-municipales-madrid}), CC BY 4.0
(\url{https://creativecommons.org/licenses/by/4.0/}). Map created by the authors.}}\label{fig:districts}
\end{figure}

\begin{figure}
\centering
\includegraphics[width=350pt]{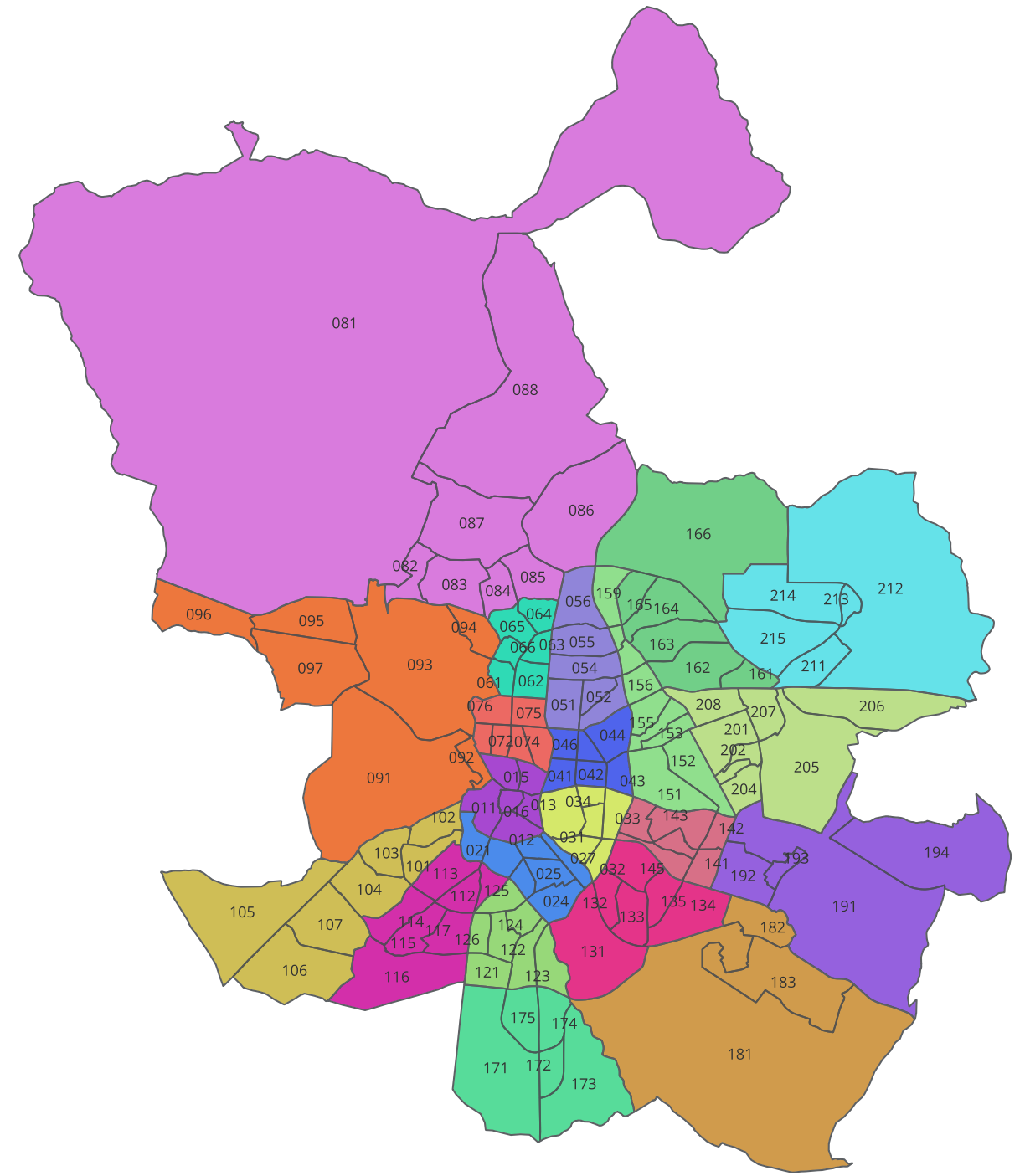}
\caption{Madrid map and their 131 neighbourhoods as of April 2025, coloured by district. \textcolor{olive}{Base layer: Ayuntamiento de Madrid, ``Barrios municipales de Madrid''
(\url{https://datos.madrid.es/dataset/300496-0-barrios-madrid}), CC BY 4.0
(\url{https://creativecommons.org/licenses/by/4.0/}). Map created by the authors.}}\label{fig:barrios}
\end{figure}

In addition, we use several official shapefiles containing the geographical limits of the various administrative entities examined. All are available in \cite{geoportalmadrid}. The city of Madrid is divided into $21$ districts, each uniquely identified with a number form $01$ to $21$; see Figure \ref{fig:districts}. Each of them is subdivided into $3$ to $9$ neighbourhoods, which, as of April 2025, total $131$; see Figure \ref{fig:barrios}. Each neighbourhood is assigned a three-digit number whose first two digits correspond to the district code. We refer to \S \ref{section:data processing} for details on the changes the neigbourhoods underwent in the last two decades.

\subsection{Data preprocessing}\label{section:data processing}

In this paper, we work with data from 2004 to 2023. During this period, the following neighbourhoods were officially created in 2017 and therefore appear in the dataset only from that year onward: Ensanche de Vallecas (district Villa de Vallecas), Valderrivas and El Cañaveral (district Vicálvaro). All of them were created as a result of two city council votings, see \cite{plenoensanche} and \cite{plenovicalvaro}.

In 2017, district `19 Vicálvaro' underwent several changes as a result of a new division into neighbourhoods \cite{plenovicalvaro}.
Neigbourhoods `193 Valderrivas' and `194 El Cañaveral' were incorporated to the district, covering geographical areas that used to be part of neighbourhood `191 Casco Histórico de Vicálvaro'.
Since these neighbourhoods were created in 2017, there is no emigration data available for these prior to this moment, in spite of the areas being populated before that date. Indeed, El Cañaveral was a ``slum" that was torn down at the beginning of the 21st century and subsequently replaced by new urban developments. This means that people lived in this area long before its official incorporation as a separate neighbourhood, but where counted as part of neighbourhood `191 Casco Histórico de Vicálvaro'.
Neighbourhood 192 is a different case. Until 2017 the number 192 corresponded to neighbourhood Ambroz, but also in 2017 Ambroz was removed and incorporated into `191 Casco Histórico de Vicálvaro'. 
Number 192 was then assigned to a new neighbourhood, called Valdebernardo. These changes must also be considered when quantifying the people who moved to/from this district.

At the turn of the century, the city of Madrid also expanded toward the South, resulting in an important growth of the district Villa de Vallecas. New housing developments were built in areas belonging to the Casco Histórico de Vallecas neighbourhood which were previously empty. Although the first inhabitants of these buildings moved there already in 2006, the new `183 Ensanche de Vallecas' neighbourhood (literally `Expansion of Vallecas') was scinded from Casco Histórico de Vallecas and officially recognised as an administrative independent neighbourhood in 2017 \cite{plenoensanche}. This is the reason why there is no data available for this neighbourhood until 2017.

The shapefile available on the Madrid city council website depicts the latest boundaries of the neighbourhoods. Thus, neighbourhoods that emerged later on as a result of splitting existing ones appear to have no emigration or inmigration until that moment, creating the false impression that they were empty until then. Moreover, these neighbourhoods are located well inside the city and not at the border, and leaving them blank until 2017 would severely interfere with the topology of the 3D cubical complexes we built.

As a consequence of the changes undergone in the neighbourhoods of these two districts, we perform the following modifications to the tabular data:
\begin{itemize}
\item From 2004 to 2016, we consider the whole district instead of individual neighbourhoods. We do so by assigning to each group we have split people moving into the same values to all the neighbourhoods in Vic\'alvaro (191, 192, 193 and 194), namely that obtained by merging 191 and 192 together. That is, someone moving within this district is considered to be staying in the same neighbourhood.
By doing so, we ensure that no `holes' are created in the map in the areas where 193 and 194 are located, as well as that the change in location for 192 is correctly taken into account.
\item From 2017 on, we keep data as is for the four neighbourhoods in district 19.
\item We follow a similar approach to correctly deal with the creation of neighbourhood 183 Ensanche de Vallecas. As its area belonged to neighbourhood 181 Casco Histórico de Vallecas between 2004 and 2016, we assigning to it the emigration rates of 181 Casco Histórico de Vallecas during this period.
\end{itemize}

\begin{figure}
\centering
\begin{subfigure}{.5\textwidth}
  \centering
  \includegraphics[width=\linewidth]{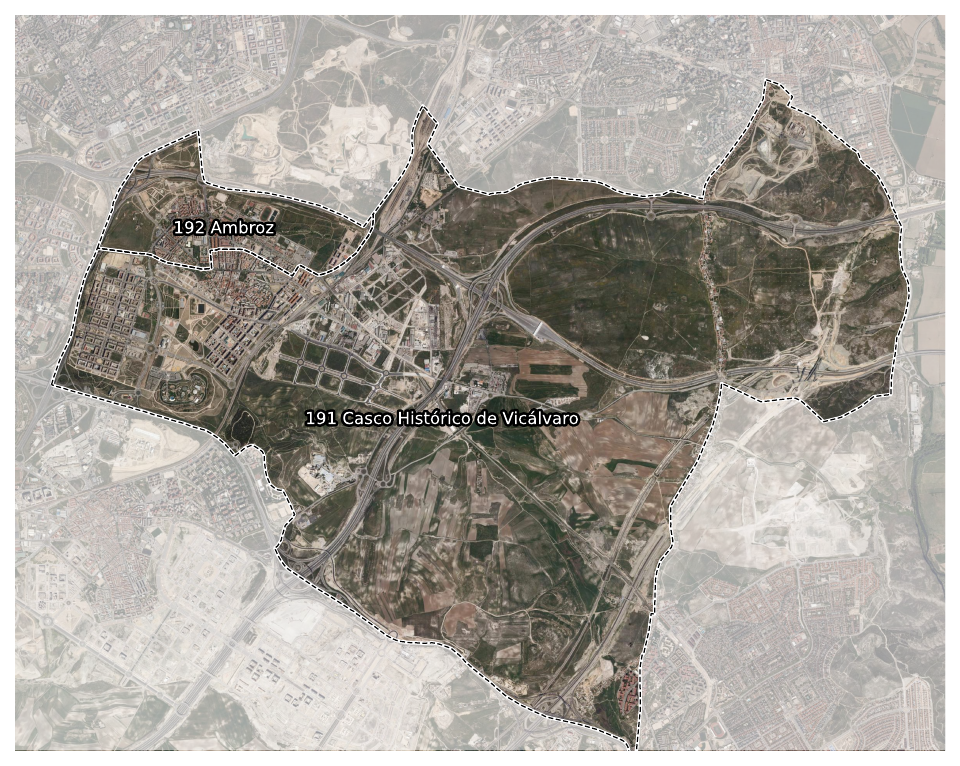}
  \caption{Until 2017, Vicálvaro consisted of two neighbourhoods: `191 Casco Histórico de Vicálvaro' and `192 Ambroz'. Background: 2006 orthophoto.}
  \label{fig:vicalvaro_before}
\end{subfigure}
\vskip\baselineskip
\begin{subfigure}{.6\textwidth}
  \centering
  \includegraphics[width=\linewidth]{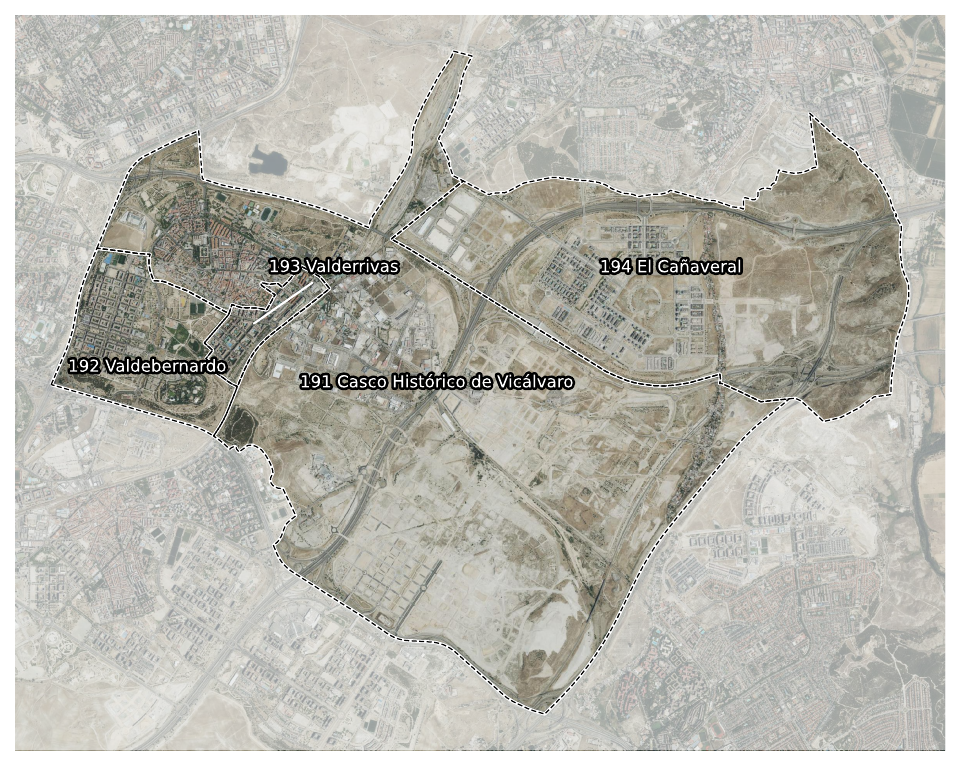}
  \caption{From 2017 on, Vicálvaro consists of four neighbourhoods: `191 Casco Histórico de Vicálvaro', `192 Valdebernardo', `193 Valderrivas' and `194 El Cañaveral' \cite{plenovicalvaro}. Notice that number 192 no longer corresponds to Ambroz but to Valdebernardo. Background: 2023 orthophoto.}
  \label{fig:vicalvaro_now}
\end{subfigure}
\caption{Changes in the neighbourhoods of district `19 Vicálvaro' in 2017. The district boundary remained essentially unchanged, but its subdivision into neighbourhoods changed. The 2006 orthophoto shows that a large part of the district had not yet been developed.
Neighbourhood boundaries: Ayuntamiento de Madrid, ``Barrios municipales de Madrid'' (\url{https://datos.madrid.es/dataset/300496-0-barrios-madrid}), historical 1987 division in (a) and current division in (b), CC BY 4.0 (\url{https://creativecommons.org/licenses/by/4.0/}).
Orthophotos: PNOA 2006 and PNOA 2023, courtesy of \textcopyright{} Instituto Geográfico Nacional (\url{https://www.ign.es/wms/pnoa-historico}), CC BY 4.0 (\url{https://www.scne.es}). Maps created by the authors.}
\label{fig:changes_neighbourhoods_vicalvaro}
\end{figure}

\begin{figure}
\centering
\begin{subfigure}{.5\textwidth}
  \centering
  \includegraphics[width=\linewidth]{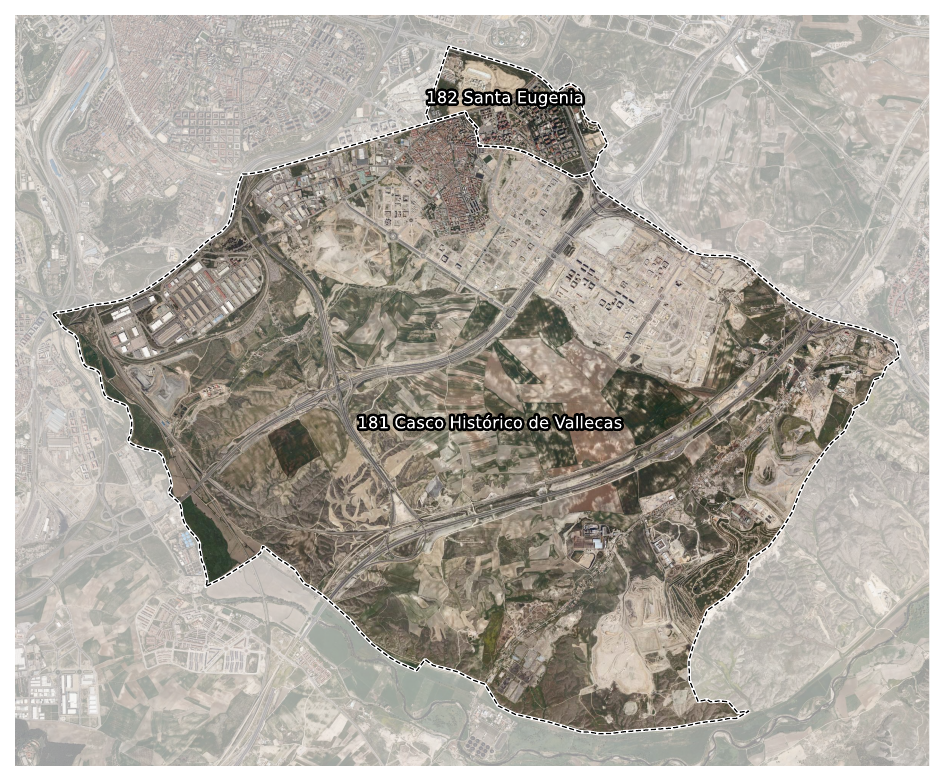}
  \caption{Until 2017, Villa de Vallecas consisted of two neighbourhoods: `181 Casco Histórico de Vallecas' and `182 Santa Eugenia'. Background: 2006 orthophoto.}
  \label{fig:vallecas_before}
\end{subfigure}
\vskip\baselineskip
\begin{subfigure}{.6\textwidth}
  \centering
  \includegraphics[width=\linewidth]{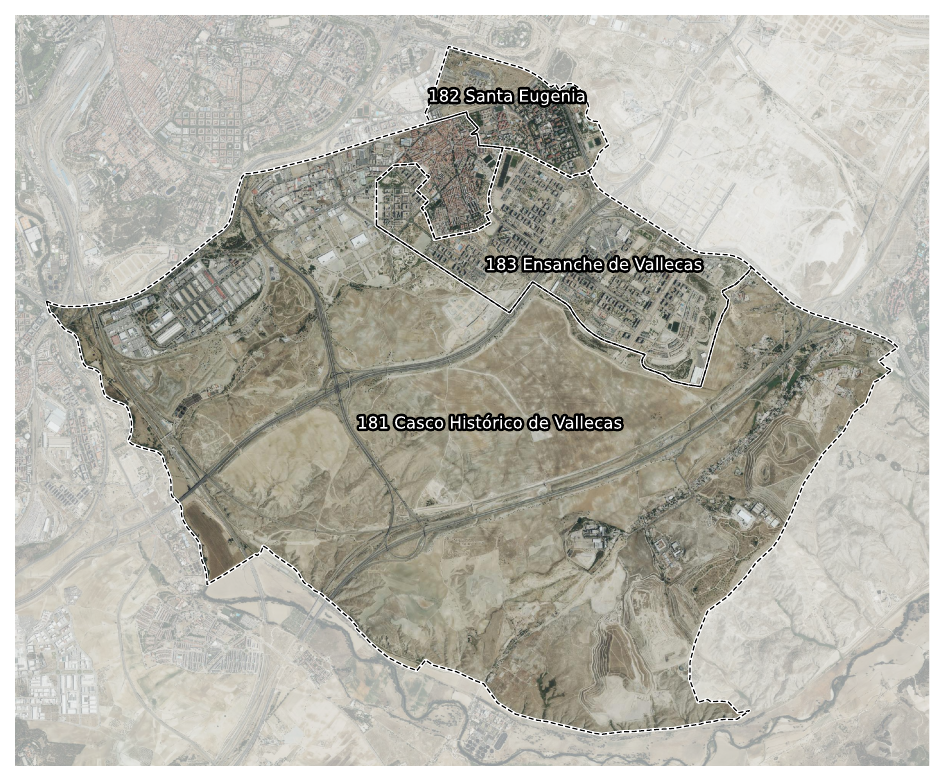}
  \caption{In 2017, part of the area formerly covered by `181 Casco Histórico de Vallecas' became the new neighbourhood `183 Ensanche de Vallecas' \cite{plenoensanche}. Background: 2023 orthophoto.}
  \label{fig:vallecas_now}
\end{subfigure}
\caption{Changes in the neighbourhoods of district `18 Villa de Vallecas' in 2017. The district boundary remained unchanged, but its subdivision into neighbourhoods changed. The 2006 orthophoto shows that a large part of the district, in particular the area of the future Ensanche de Vallecas, had not yet been developed.
Neighbourhood boundaries: Ayuntamiento de Madrid, ``Barrios municipales de Madrid'' (\url{https://datos.madrid.es/dataset/300496-0-barrios-madrid}), historical 1987 division in (a) and current division in (b), CC BY 4.0 (\url{https://creativecommons.org/licenses/by/4.0/}).
Orthophotos: PNOA 2006 and PNOA 2023, courtesy of \textcopyright{} Instituto Geográfico Nacional (\url{https://www.ign.es/wms/pnoa-historico}), CC BY 4.0 (\url{https://www.scne.es}). Maps created by the authors.}
\label{fig:changes_neighbourhoods_vallecas}
\end{figure}

\section{Cubical complexes}{\label{app:cubical}}

A $d$-dimensional grayscale digital image of size $(n_1, n_2, \ldots , n_d)$ is
a real-valued function $\mathcal{I}$ defined on a $d$--dimensional rectangular grid
\[I = \llbracket 1, n_1 \rrbracket \times  \llbracket 1, n_2 \rrbracket \times \cdots \times \llbracket 1, n_d \rrbracket  \to \mathbb{R},\]
where $\llbracket 1, n_i \rrbracket$ is the set $\{ k \in \mathbb{N} | 1 \leq k \leq n_i \}$.

Building on the analogy with digital photography, an element $p \in I$ called a pixel if $d=2$ and a voxel if $d \geq 3$.
Notice that with this construction voxels are also connected diagonally.


A cubical complex $X \subset \mathbb{R}^d$ (see \cite{kaczynski_computational, strombom_cubical})
is a cell complex consisting of a set of products of $d$ intervals
$$\sigma = e_1 \times e_2 \times \ldots \times e_d,$$
where $e_i$ may be of the form $e_i = [l_i, l_i + 1]$ or of the form $e_i = [l_i, l_i]$ with $l_i \in \mathbb{Z}$
and such that all faces of $\sigma \in X$ are also in $X$.

There are two common ways to build a filtered cell complex from a grayscale image: the T-- and the V--constructions. In the following, we recall the T--construction, which is the one implemented in Cubical Ripser \cite{cubicalripser} and used in this study.
It follows a top-down approach to building a filtered cell complex.
We refer to \cite{Bleile_dual} for a comprehensive review of both constructions and for a series of duality results between them.

Given a $d$--dimensional grayscale digital image $\mathcal{I} : I \to \mathbb{R}$ of size $(n_1, n_2, . . . , n_d)$, 
the T-construction is the filtered cell complex $(X, V)$ defined as:
\begin{enumerate}
\item $X$ is a cubical complex built from the array of $n_1 \times \cdots \times n_d$.
\item The $d$--cells $\tau^d \in X$  correspond exactly to the elements $p \in I$.
We define the function $V$ on $X$ by extending $\mathcal{I}$ to lower-dimensional cells $\sigma$ by setting
$$V(\sigma) = \min_{\sigma \preceq  \tau^d} (V(\tau^d)).$$
In other words, on a $k$--cube $V$ takes the smallest value of all adjacent top-dimensional cubes.
\end{enumerate}

In other words, each pixel is assigned a parameter value and is included in the filtration precisely when the filtration parameter attains that threshold. When a pixel is added to the filtration and one or more of its neighbours are already present, it adopts the minimum value of those neighbours for homological computation purposes.


\begin{figure}
\centering
\includegraphics[width=0.9\textwidth]{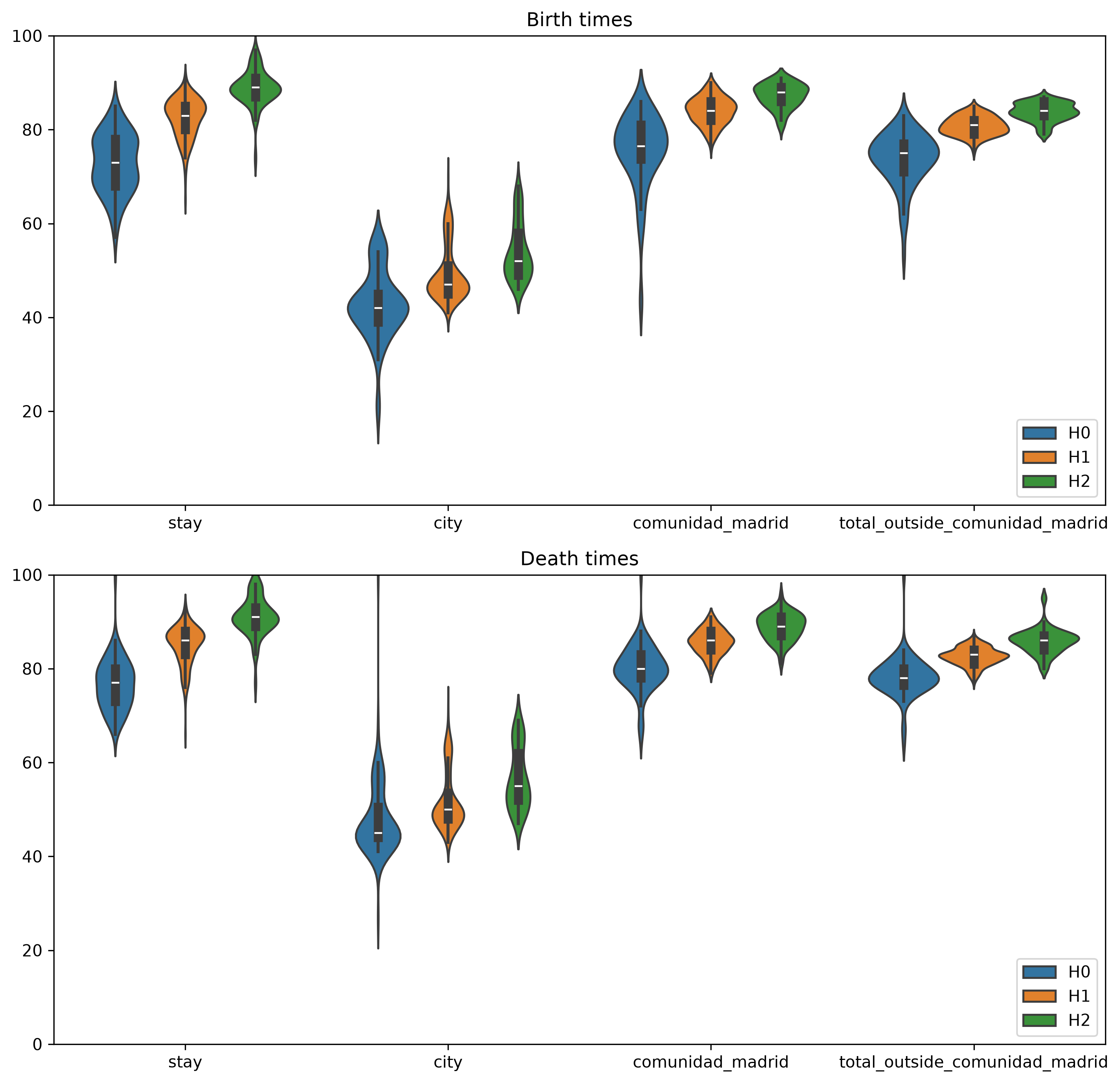}
\caption{Violinplots of the distribution of the birth and death times of the topological features of each group.}\label{fig:violinplots_birth_death}
\end{figure}


\begin{figure}
\centering
\begin{subfigure}{.38\textwidth}
  \centering
  \includegraphics[width=\linewidth]{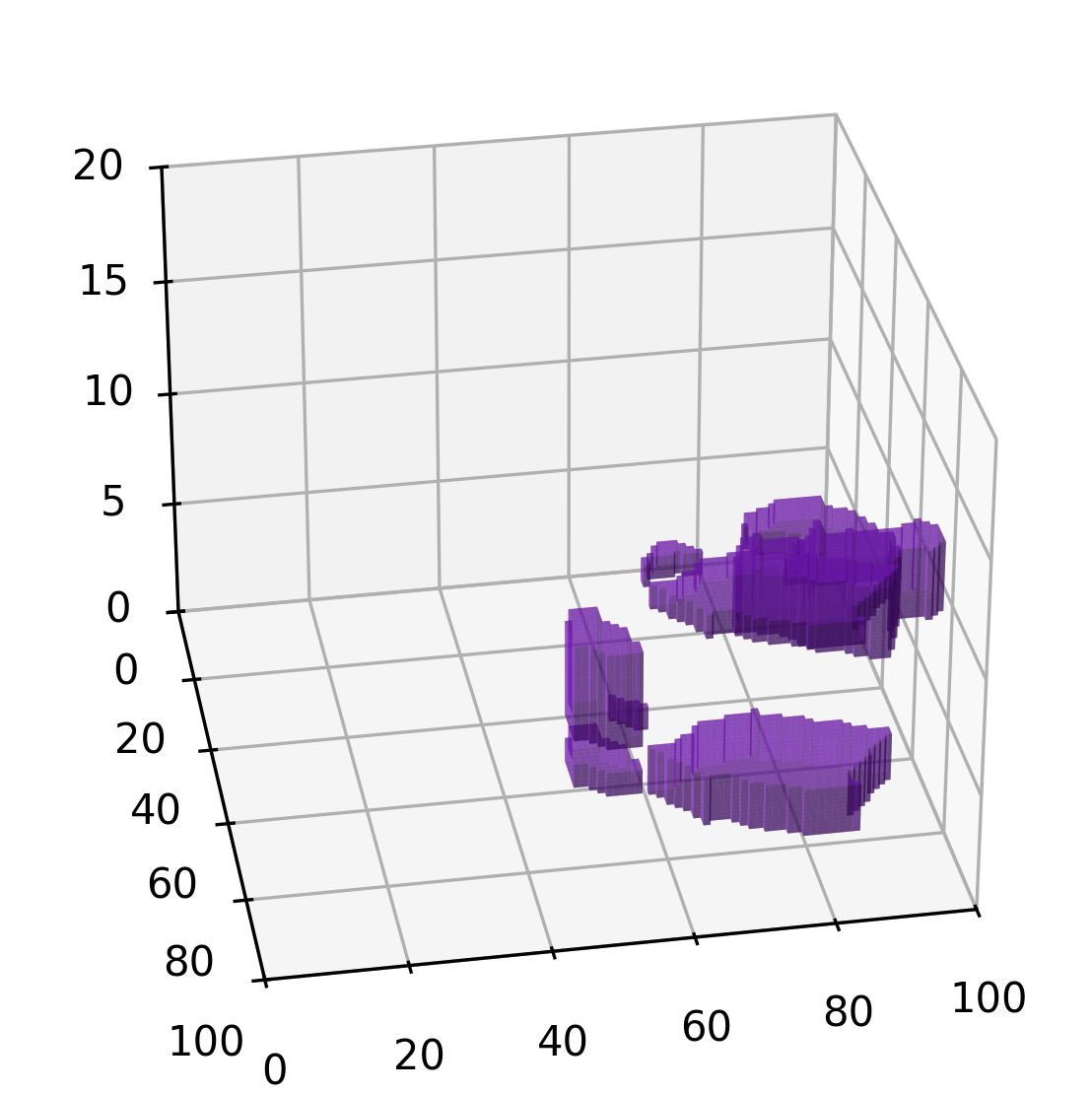}
  \label{fig:stay_threshold_35}
\end{subfigure}

\vskip\baselineskip

\begin{subfigure}{.38\textwidth}
  \centering
  \includegraphics[width=\linewidth]{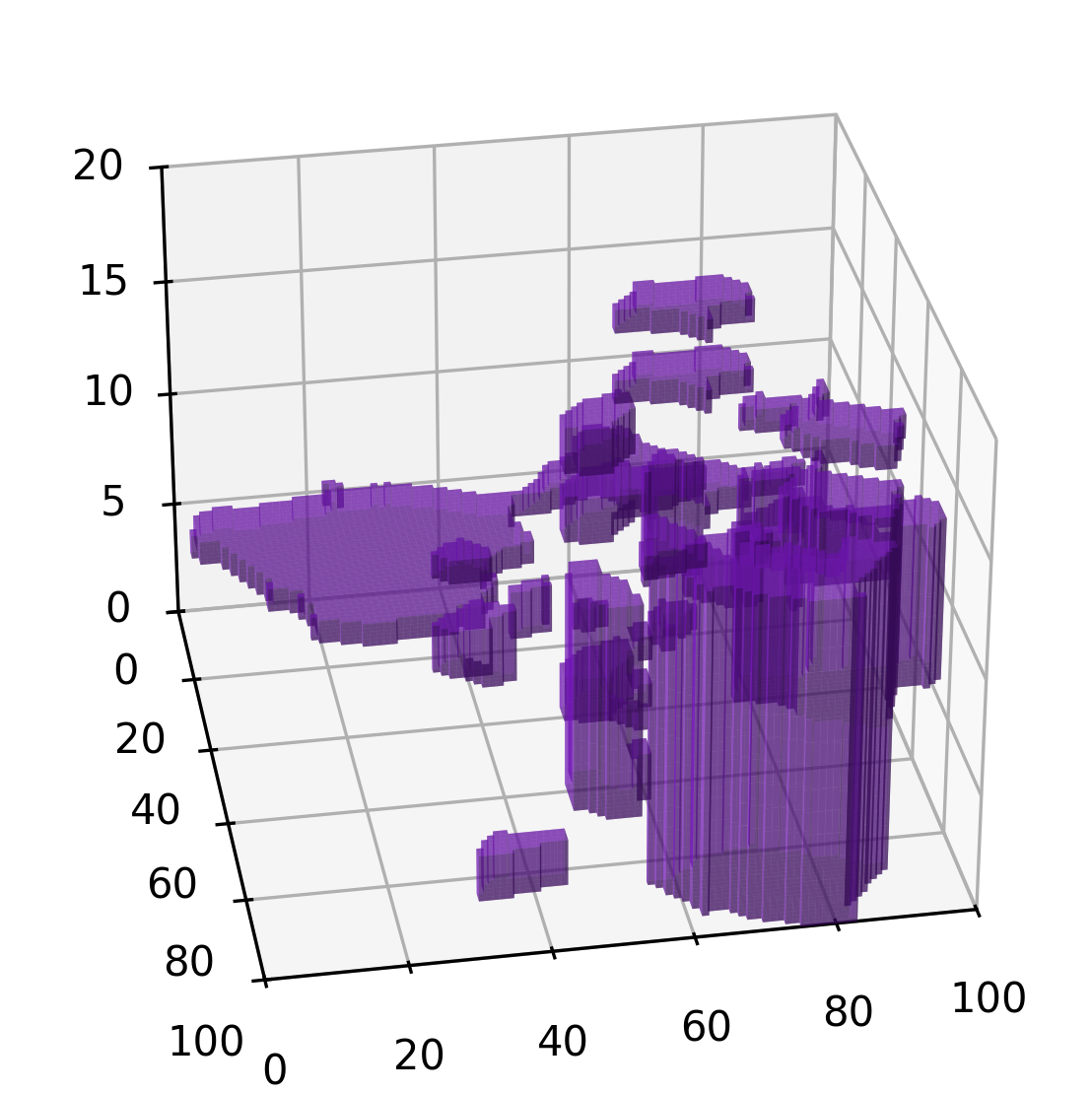}
  \label{fig:stay_threshold_30}
\end{subfigure}

\vskip\baselineskip

\begin{subfigure}{.38\textwidth}
  \centering
  \includegraphics[width=\linewidth]{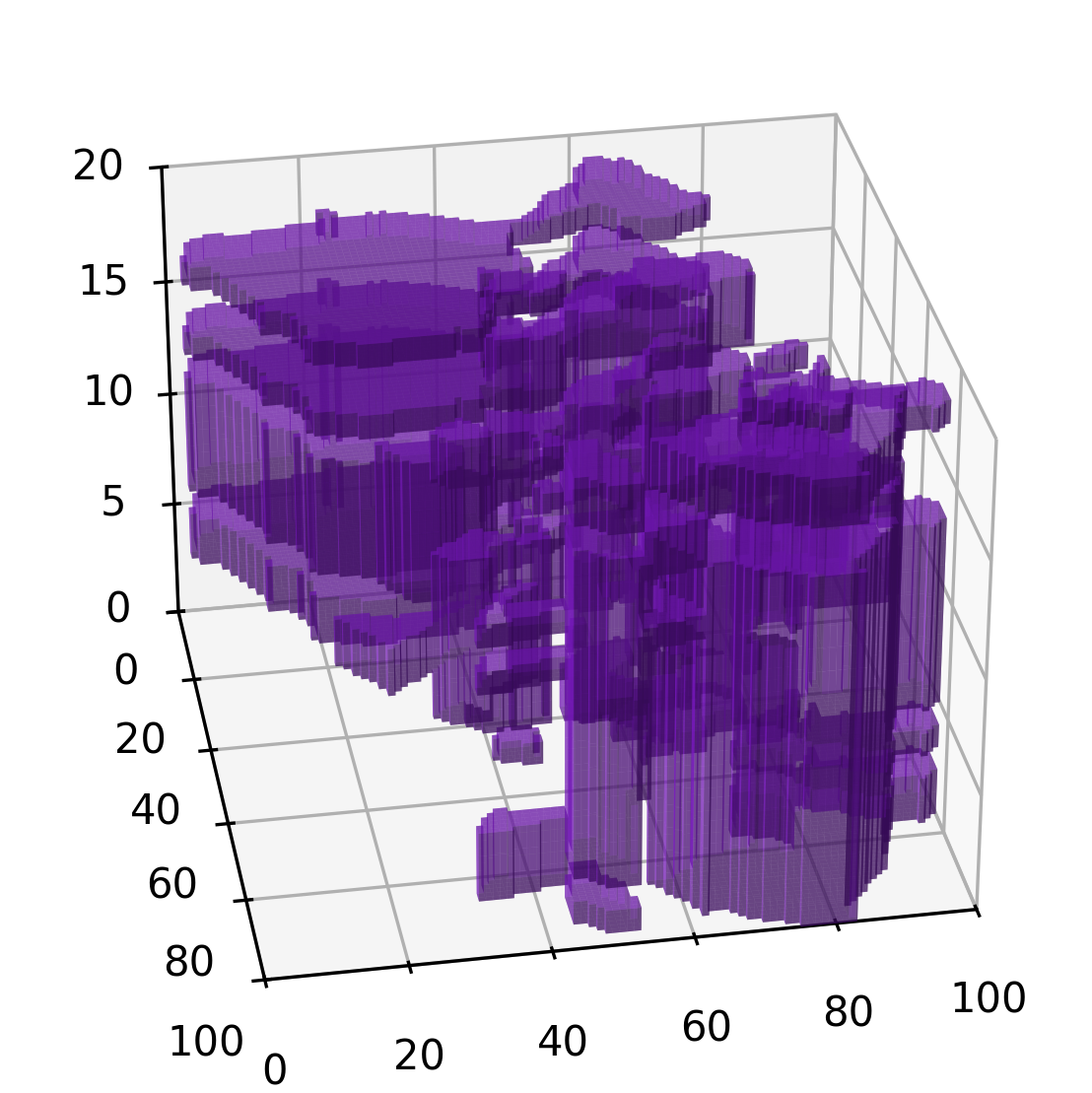}
  \label{fig:stay_threshold_25}
\end{subfigure}
\caption{Cubical complexes for the group 'stay' at filtration-parameter values 65, 70 and 75. \textcolor{olive}{Derived from neighbourhood boundaries by the Ayuntamiento de Madrid
(\url{https://datos.madrid.es/dataset/300496-0-barrios-madrid}), CC BY 4.0.}}
\label{fig:3d_complexes_stay}
\end{figure}


\begin{figure}
\centering
\begin{subfigure}{.38\textwidth}
  \centering
  \includegraphics[width=\linewidth]{fig/city_threshold_70.png}
  \label{fig:city_threshold_70}
\end{subfigure}

\vskip\baselineskip

\begin{subfigure}{.38\textwidth}
  \centering
  \includegraphics[width=\linewidth]{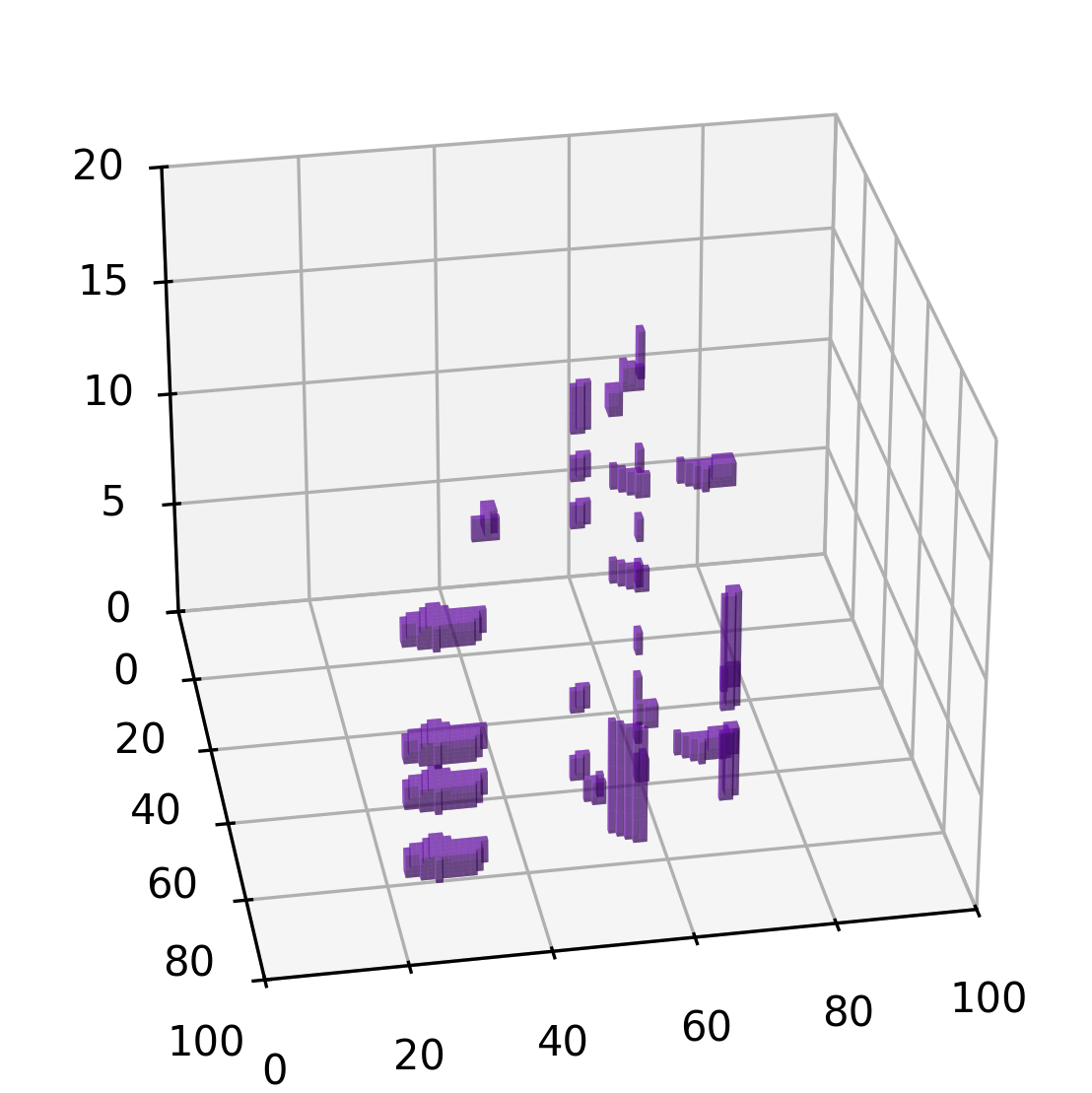}
  \label{fig:city_threshold_60}
\end{subfigure}

\vskip\baselineskip

\begin{subfigure}{.38\textwidth}
  \centering
  \includegraphics[width=\linewidth]{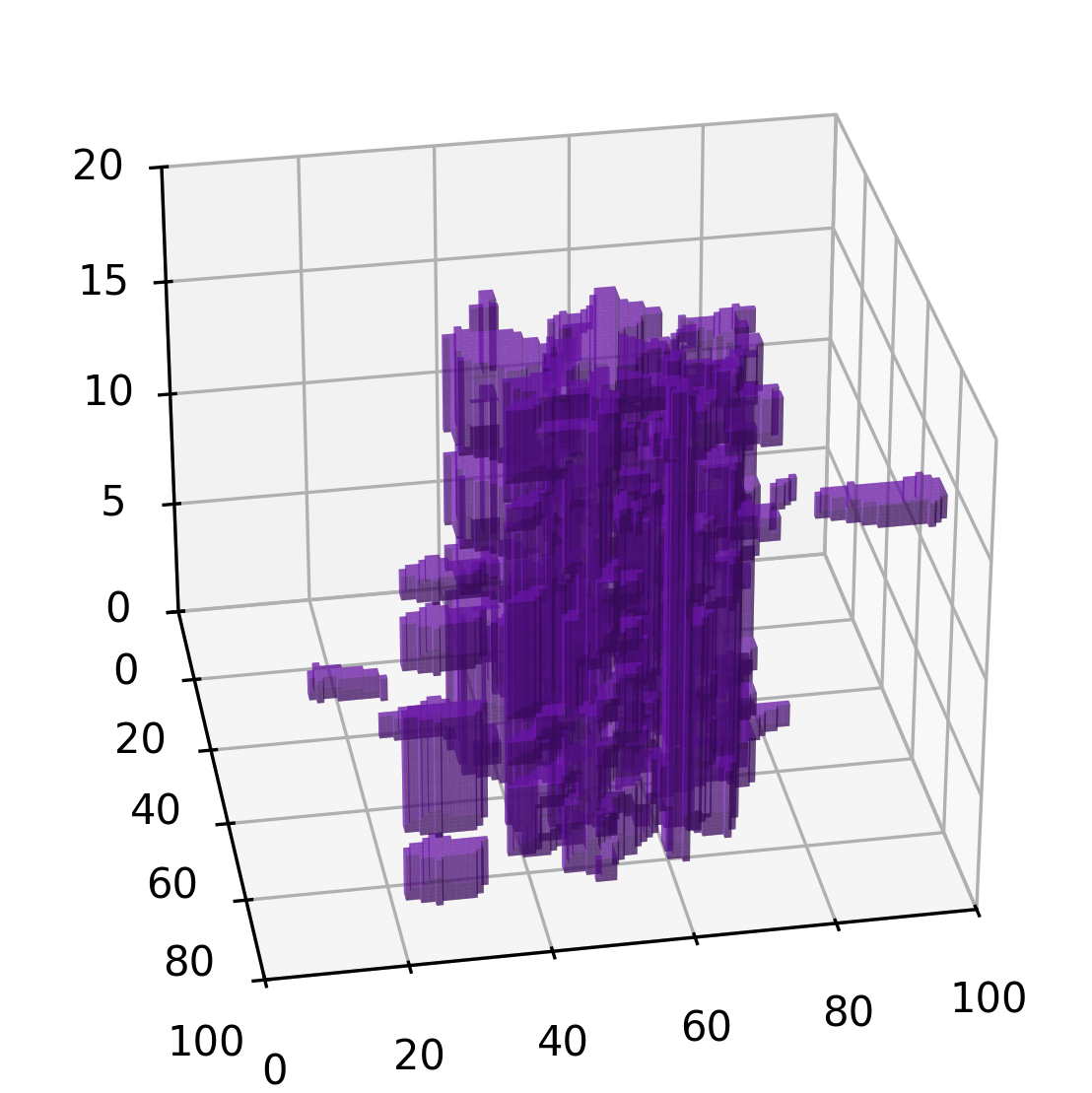}
  \label{fig:city_threshold_50}
\end{subfigure}
\caption{Cubical complexes for the group 'city' at filtration-parameter values 30, 40 and 50. \textcolor{olive}{Derived from neighbourhood boundaries by the Ayuntamiento de Madrid
(\url{https://datos.madrid.es/dataset/300496-0-barrios-madrid}), CC BY 4.0.}}
\label{fig:3d_complexes_city}
\end{figure}


\begin{figure}
\centering
\begin{subfigure}{.38\textwidth}
  \centering
  \includegraphics[width=\linewidth]{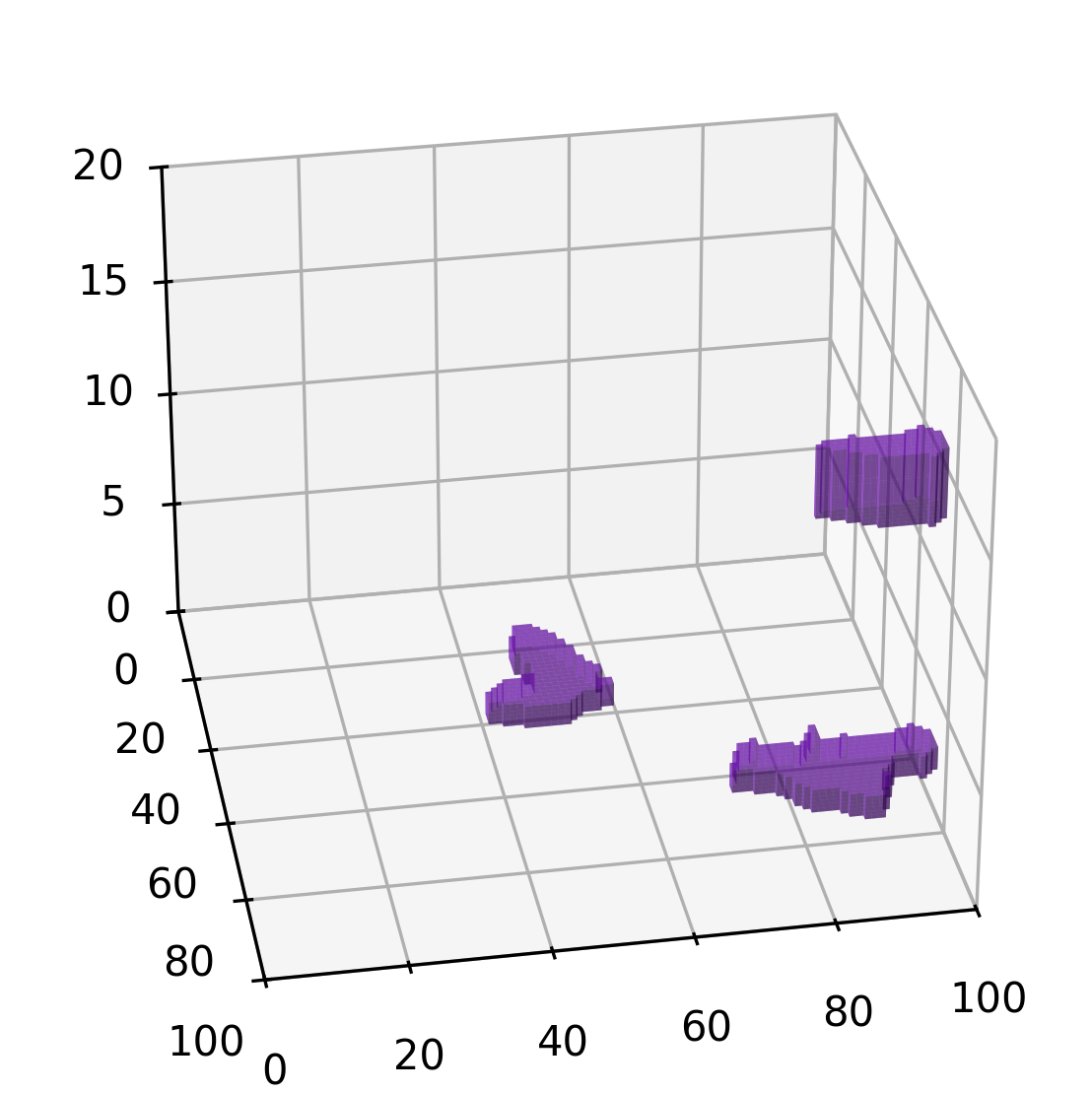}
  \label{fig:cm_threshold_40}
\end{subfigure}

\vskip\baselineskip

\begin{subfigure}{.38\textwidth}
  \centering
  \includegraphics[width=\linewidth]{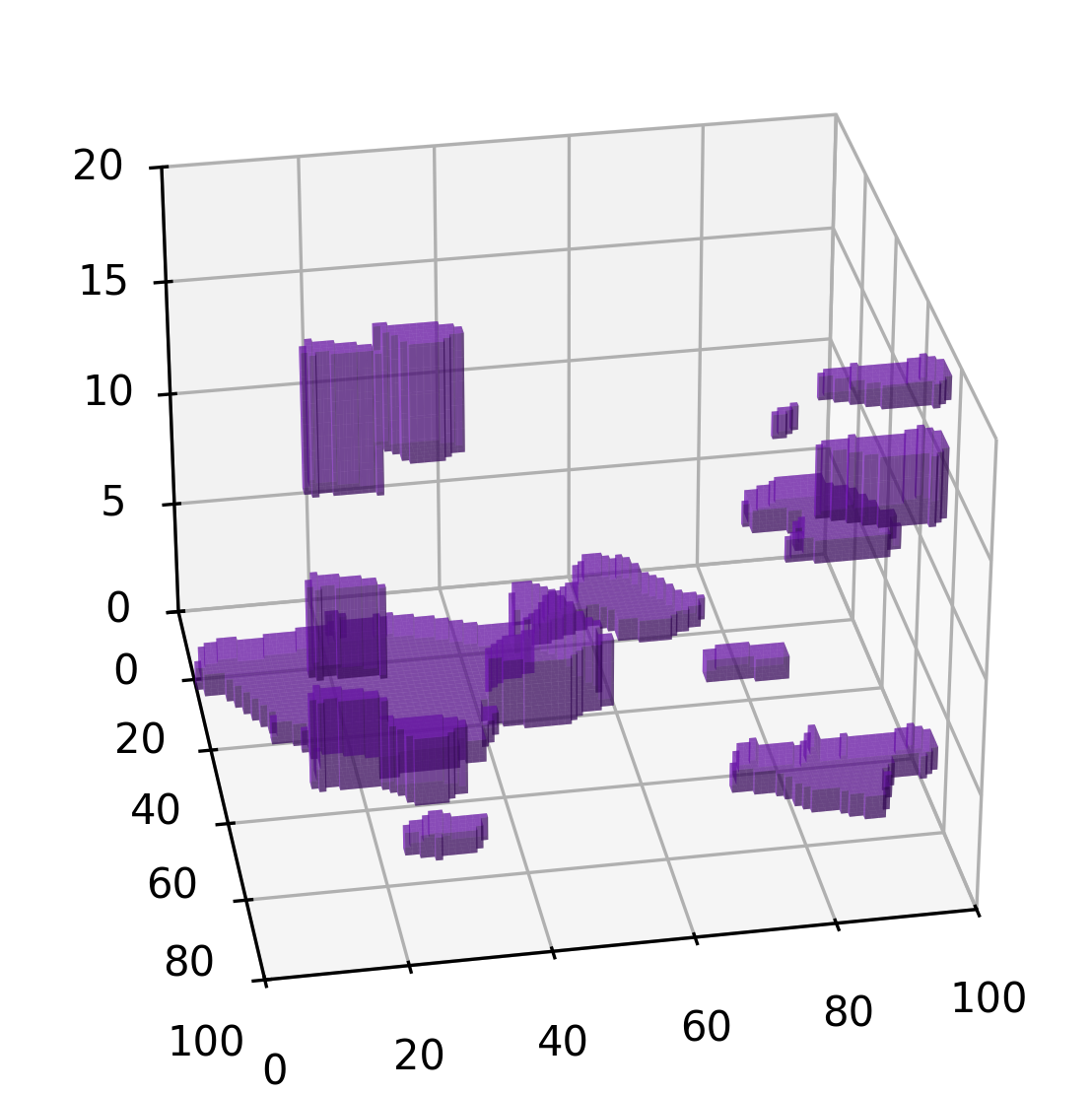}
  \label{fig:cm_threshold_30}
\end{subfigure}

\vskip\baselineskip

\begin{subfigure}{.38\textwidth}
  \centering
  \includegraphics[width=\linewidth]{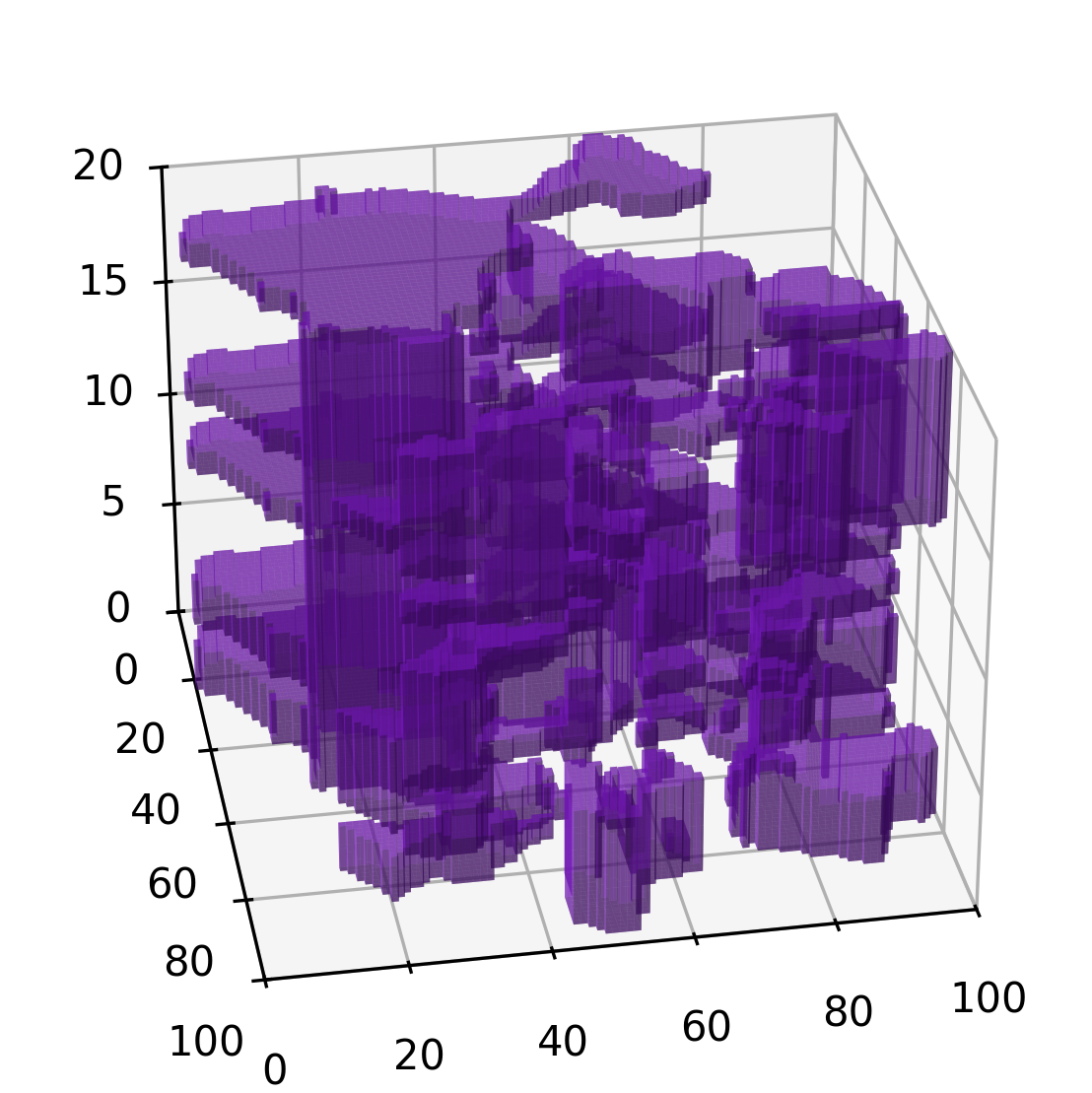}
  \label{fig:cm_threshold_20}
\end{subfigure}
\caption{Cubical complexes for the group 'Comunidad de Madrid' at filtration-parameter values 60, 70 and 80. \textcolor{olive}{Derived from neighbourhood boundaries by the Ayuntamiento de Madrid
(\url{https://datos.madrid.es/dataset/300496-0-barrios-madrid}), CC BY 4.0.}}
\label{fig:3d_complexes_cm}
\end{figure}


\begin{figure}
\centering
\begin{subfigure}{.38\textwidth}
  \centering
  \includegraphics[width=\linewidth]{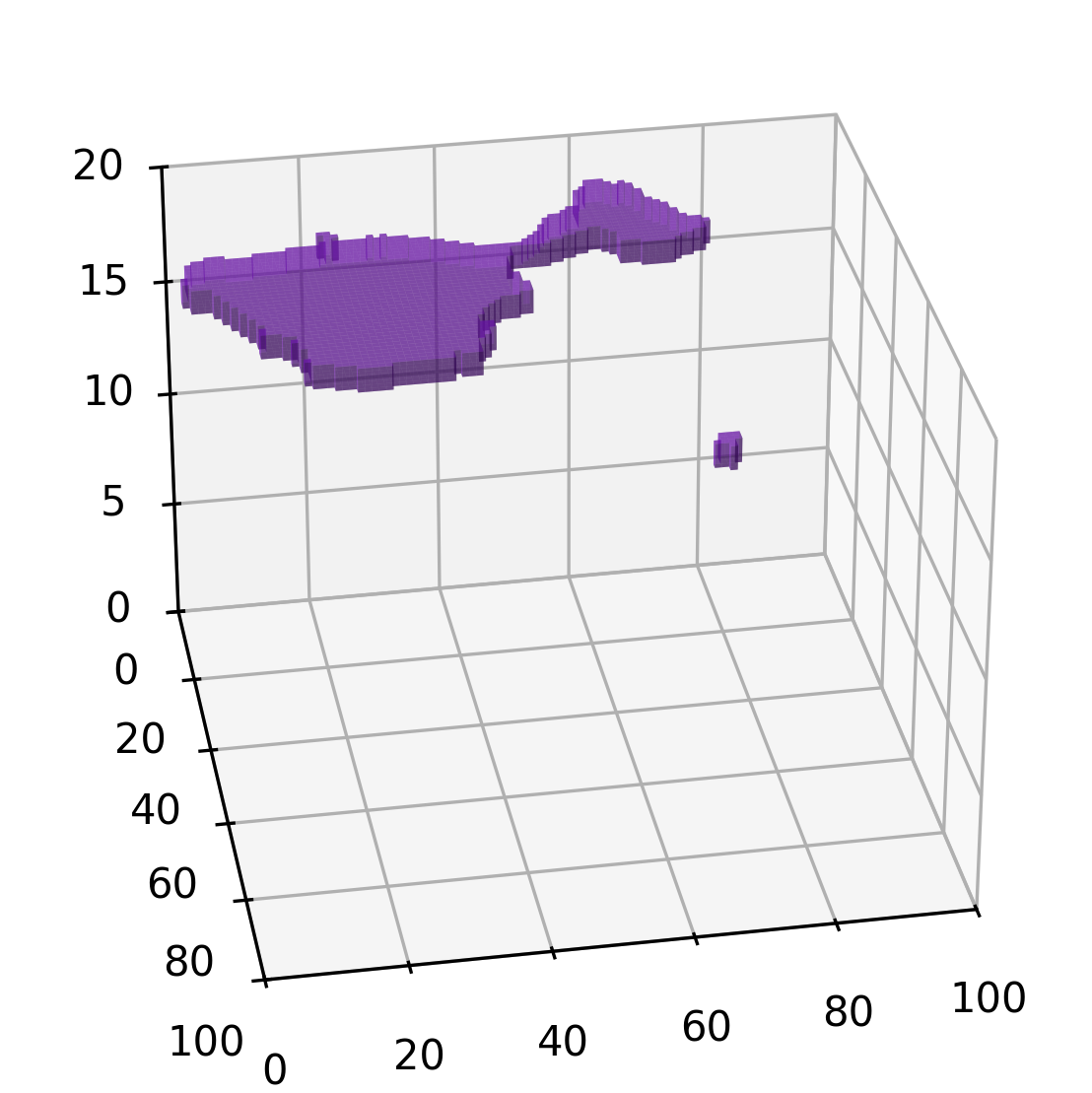}
  \label{fig:outside_threshold_40}
\end{subfigure}

\vskip\baselineskip

\begin{subfigure}{.38\textwidth}
  \centering
  \includegraphics[width=\linewidth]{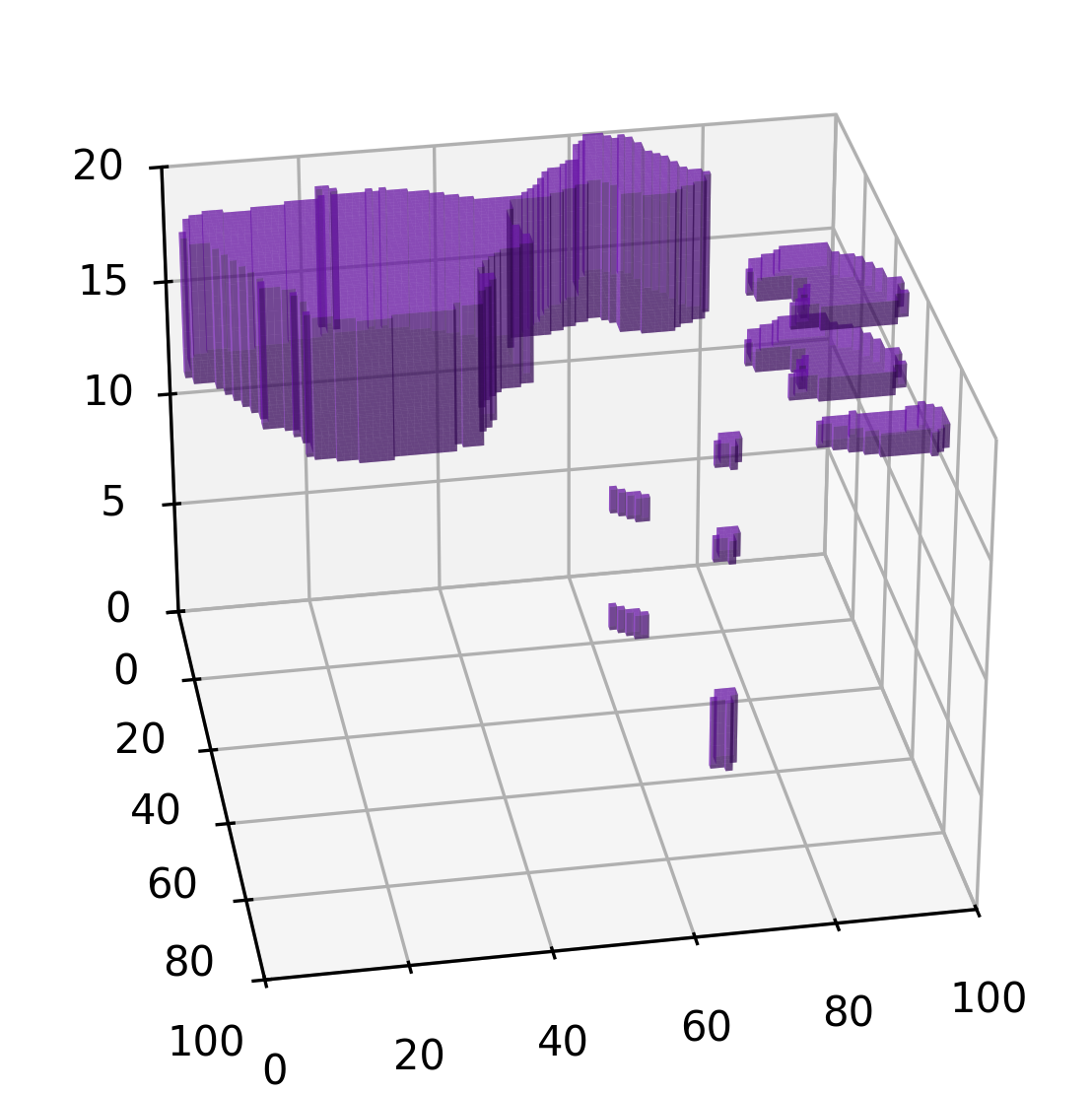}
  \label{fig:outside_threshold_30}
\end{subfigure}

\vskip\baselineskip

\begin{subfigure}{.38\textwidth}
  \centering
  \includegraphics[width=\linewidth]{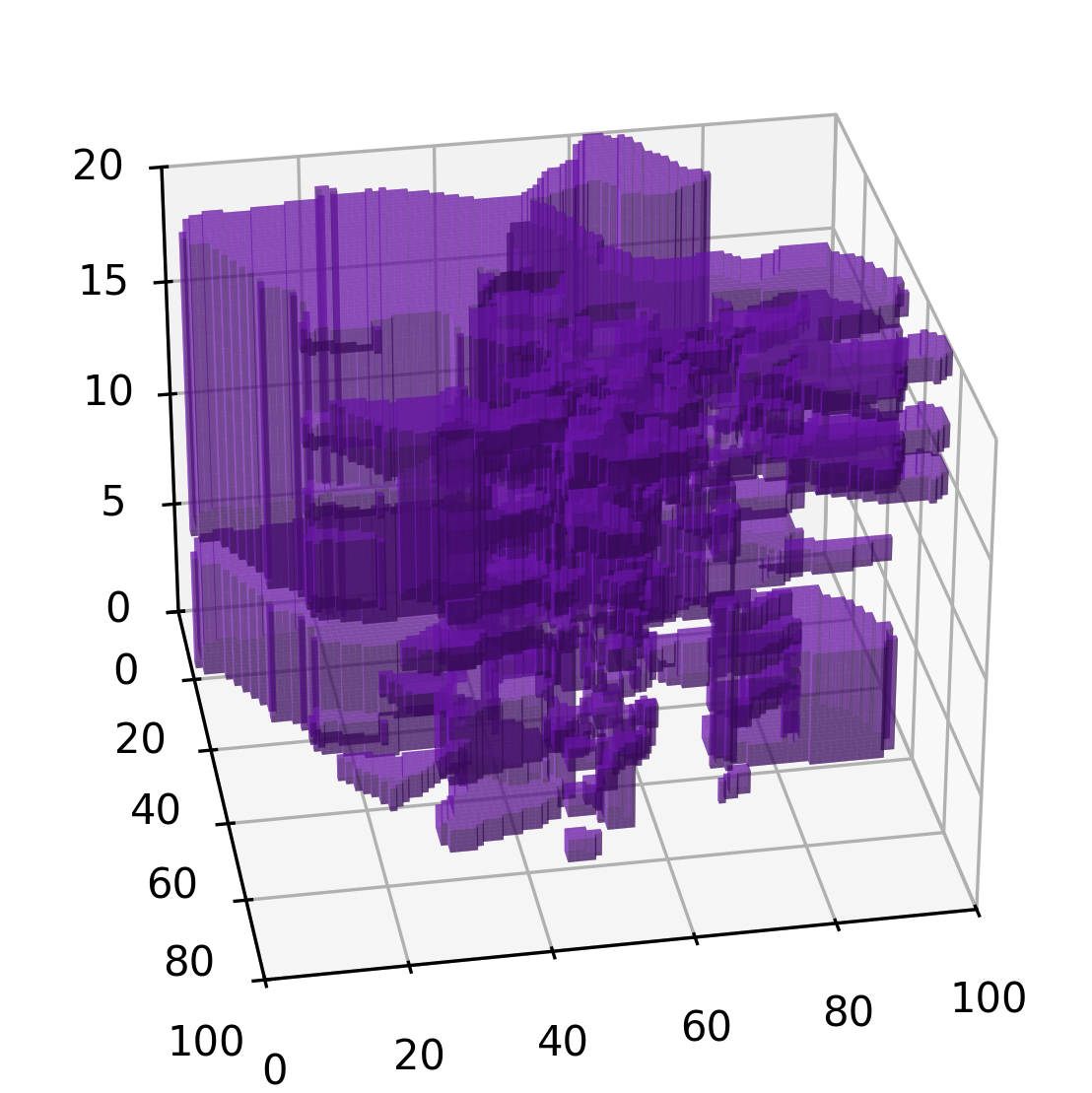}
  \label{fig:outside_threshold_20}
\end{subfigure}
\caption{Cubical complexes for the group 'outside' at filtration-parameter values 60, 70 and 80. \textcolor{olive}{Derived from neighbourhood boundaries by the Ayuntamiento de Madrid
(\url{https://datos.madrid.es/dataset/300496-0-barrios-madrid}), CC BY 4.0.}}
\label{fig:3d_complexes_outside}
\end{figure}


\begin{table}[t!]
\centering
\begin{tabular}{lllrrrr}
 & dim & measure & stay & city & C. Madrid & outside \\
\midrule
birth & H0 & mean & 72.79 & 42.68 & 75.45 & 73.64 \\
 & H0 & median & 73.0 & 42.0 & 76.5 & 75.0 \\
 & H0 & std & 6.20 & 7.27 & 8.15 & 5.66 \\
 & H0 & min & 57.0 & 19.0 & 43.0 & 53.0 \\ 
 & H0 & max & 85.0 & 57.0 & 86.0 & 83.0 \\
 & H1 & mean & 82.53 & 49.31 & 84.04 & 80.80 \\
 & H1 & median & 83.0 & 47.0 & 84.0 & 81.0 \\
 & H1 & std & 4.19 & 5.82 & 2.93 & 2.02 \\
 & H1 & min & 65.0 & 41.0 & 76.0 & 75.0 \\
 & H1 & max & 91.0 & 70.0 & 90.0 & 85.0 \\
 & H2 & mean &  88.84 & 54.05 & 87.24 & 83.74 \\
 & H2 & median & 89.0 & 52.0 & 88.0 & 84.0 \\
 & H2 & std & 3.72 & 6.07 & 2.61 & 2.01 \\
 & H2 & min & 73.0 & 46.0 & 80.0 & 79.0 \\
 & H2 & max & 97.0 & 68.0 & 91.0 & 87.0 \\
\midrule
death & H0 & mean & 76.63 & 47.79 & 80.01 & 78.13 \\
 & H0 & median & 77.0 & 45.0 & 80.0 & 78.0 \\
 & H0 & std & 5.466 & 8.23 & 4.94 & 4.28 \\
 & H0 & min & 66.0 & 27.0 & 65.0 & 64.0 \\
 & H0 & max & 100.0 & 100.0 & 100.0 & 100.0 \\
 & H1 & mean & 84.96 & 51.70 & 85.66 & 82.57 \\
 & H1 & median & 86.0 & 50.0 & 86.0 & 83.0 \\
 & H1 & std & 4.10 & 6.07 & 2.67 & 1.92 \\
 & H1 & min & 66.0 & 43.0 & 79.0 & 77.0 \\
 & H1 & max & 93.0 & 72.0 & 91.0 & 87.0 \\
 & H2 & mean &  91.09 & 56.59 & 88.91 & 85.80 \\
 & H2 & median & 91.0 & 55.0 & 89.0 & 86.0 \\
 & H2 & std & 4.06 & 6.54 & 2.87 & 2.72 \\
 & H2 & min & 76.0 & 47.0 & 81.0 & 80.0 \\
 & H2 & max & 100.0 & 69.0 & 96.0 & 95.0 \\
\midrule
persistence & H0 & mean & 3.83 & 5.10 & 4.565 & 4.49 \\
 & H0 & median & 2.0 & 3.0 & 2.0 & 3.0 \\
 & H0 & std & 5.51 & 8.93 & 7.89 & 6.36 \\
 & H0 & min & 1.0 & 1.0 & 1.0 & 1.0 \\
 & H0 & max & 43.0 & 81.0 & 57.0 & 47.0 \\
 & H1 & mean & 2.426 & 2.39 & 1.63 & 1.77 \\
 & H1 & median & 2.0 & 2.0 & 1.0 & 1.0 \\
 & H1 & std & 1.97 & 1.48 & 1.07 & 0.9 \\
 & H1 & min & 1.0 & 1.0 & 1.0 & 1.0 \\
 & H1 & max & 15.0 & 8.0 & 7.0 & 5.0 \\
 & H2 & mean & 2.25 & 2.54 & 1.68 & 2.06 \\
 & H2 & median & 2.0 & 2.0 & 1.0 & 1.0 \\
 & H2 & std & 1.57 & 2.53 & 1.32 & 1.92 \\
 & H2 & min & 1.0 & 1.0 & 1.0 & 1.0 \\
 & H2 & max & 9.0 & 16.0 & 9.0 & 12.0 \\
\bottomrule
\end{tabular}
\caption{Descriptive statistics of the topological features obtained, by group and dimension.}\label{table:stats}
\end{table}

\begin{table}\centering
\resizebox{!}{0.4\textwidth}{\begin{tabular}{lrrrrrrr}
dim&birth&death&persistence&year of birth&year of death&neighbourhood of birth&neighbourhood of death\\
\midrule
0 & 57 & 100 & 43 & 2009 &  & 27 & \\
0 & 60 & 74 & 14 & 2013 & 2011 & 171 & 132 \\
0 & 64 & 78 & 14 & 2012 & 2020 & 212 & 107 \\
0 & 65 & 75 & 10 & 2010 & 2012 & 81 & 81 \\
0 & 67 & 77 & 10 & 2005 & 2008 & 86 & 86 \\
0 & 61 & 70 & 9 & 2009 & 2014 & 181 & 131 \\
0 & 64 & 73 & 9 & 2012 & 2015 & 164 & 212 \\
0 & 71 & 80 & 9 & 2004 & 2006 & 83 & 87 \\
0 & 66 & 73 & 7 & 2005 & 2007 & 116 & 171 \\
0 & 62 & 68 & 6 & 2013 & 2011 & 194 & 181 \\
1 & 78 & 93 & 15 & 2016 & 2019 & 205 & 158 \\
1 & 77 & 86 & 9 & 2023 & 2022 & 95 & 95 \\
1 & 78 & 87 & 9 & 2007 & 2008 & 164 & 166 \\
1 & 69 & 77 & 8 & 2021 & 2020 & 191 & 191 \\
1 & 78 & 86 & 8 & 2013 & 2013 & 201 & 203 \\
1 & 79 & 87 & 8 & 2022 & 2020 & 159 & 159 \\
1 & 81 & 89 & 8 & 2014 & 2013 & 62 & 16 \\
1 & 72 & 79 & 7 & 2018 & 2018 & 181 & 174 \\
1 & 78 & 85 & 7 & 2015 & 2010 & 125 & 25 \\
1 & 79 & 86 & 7 & 2016 & 2015 & 24 & 122 \\
2 & 89 & 98 & 9 & 2018 & 2019 & 25 & 27 \\
2 &	89 & 97	& 8 & 2018 & 2015 &	25 & 27 \\
2 & 90 & 97 & 7 & 2017 & 2015 & 13 & 27 \\
2 &	90 & 97 & 7 & 2016 & 2015 &	25 & 27 \\
2 & 85 & 91 & 6 & 2022 & 2020 & 193 & 193 \\
2 & 92 & 98 & 6 & 2020 & 2019 & 13 & 27 \\
2 & 89 & 94 & 5 & 2010 & 2012 & 93 & 82 \\
2 & 93 & 98 & 5 & 2023 & 2020 & 141 & 141 \\
2 & 95 & 100 & 5 & 2007 & 2008 & 27 & 27 \\
2 &	85 & 89 & 4 & 2011 & 2011 &	162 & 215 \\
\bottomrule
\end{tabular}}
\caption{Top $10$ most persistent features in each dimension for the group `stay'}
\end{table}

\begin{table}\centering
\resizebox{!}{0.4\textwidth}{\begin{tabular}{lrrrrrrr}
dim&birth&death&persistence&year of birth&year of death&neighbourhood of birth&neighbourhood of death\\
\midrule
0 & 19 & 100 & 81 & 2008 &  & 27 & \\
0 & 22 & 45 & 23 & 2019 & 2018 & 27 & 26 \\
0 & 35 & 58 & 23 & 2008 & 2009 & 106 & 105 \\
0 & 46 & 68 & 22 & 2017 & 2016 & 194 & 191 \\
0 & 39 & 53 & 14 & 2005 & 2006 & 106 & 106 \\
0 & 31 & 44 & 13 & 2015 & 2013 & 27 & 16 \\
0 & 39 & 52 & 13 & 2015 & 2013 & 106 & 106 \\
0 & 32 & 43 & 11 & 2010 & 2006 & 142 & 43 \\
0 & 33 & 43 & 10 & 2006 & 2006 & 158 & 62 \\
0 & 34 & 44 & 10 & 2012 & 2022 & 158 & 54 \\
1 & 57 & 65 & 8 & 2010 & 2009 & 214 & 213 \\
1 & 41 & 48 & 7 & 2005 & 2007 & 35 & 31 \\
1 & 44 & 51 & 7 & 2011 & 2010 & 32 & 31 \\
1 & 57 & 64 & 7 & 2006 & 2007 & 214 & 213 \\
1 & 45 & 51 & 6 & 2013 & 2012 & 25 & 26 \\
1 & 51 & 57 & 6 & 2006 & 2007 & 163 & 164 \\
1 & 57 & 63 & 6 & 2014 & 2015 & 214 & 213 \\
1 & 62 & 68 & 6 & 2021 & 2021 & 96 & 95 \\
1 & 43 & 48 & 5 & 2006 & 2006 & 154 & 52 \\
1 & 44 & 49 & 5 & 2013 & 2013 & 66 & 51 \\

2 & 49 & 65 & 16 & 2018 & 2018 & 31 & 27 \\
2 & 54 & 65 & 11 & 2020 & 2020 & 31 & 27 \\
2 & 56 & 65 & 9 & 2020 & 2020 & 134 & 141 \\
2 & 51 & 59 & 8 & 2010 & 2009 & 31 & 27 \\
2 & 51 & 58 & 7 & 2006 & 2012 &	12 & 12 \\
2 & 58 & 65 & 7 & 2018 & 2018 & 181 & 27 \\
2 & 59 & 66 & 7 & 2023 & 2013 & 165 & 164 \\
2 & 60 & 65 & 5 & 2012 & 2013 & 205 & 207 \\
2 &	49 & 54 & 5 & 2011 & 2013 &	112 & 111 \\
2 & 48 & 52 & 4 & 2020 & 2017 & 126 & 111 \\
\bottomrule
\end{tabular}}
\caption{Top $10$ most persistent features in each dimension for the group `city'}
\end{table}

\begin{table}\centering
\resizebox{!}{0.4\textwidth}{\begin{tabular}{lrrrrrrr}
dim&birth&death&persistence&year of birth&year of death&neighbourhood of birth&neighbourhood of death\\
\midrule
0 & 43 & 100 & 57 & 2004 &  & 88 & \\
0 & 44 & 81 & 37 & 2018 & 2020 & 194 & 191 \\ 
0 & 57 & 78 & 21 & 2006 & 2010 & 194 & 86 \\
0 & 64 & 78 & 14 & 2013 & 2009 & 212 & 211 \\
0 & 60 & 72 & 12 & 2010 & 2008 & 96 & 97 \\
0 & 62 & 74 & 12 & 2020 & 2013 & 96 & 97 \\
0 & 69 & 81 & 12 & 2021 & 2020 & 193 & 191 \\
0 & 68 & 77 & 9 & 2007 & 2012 & 214 & 166 \\
0 & 69 & 78 & 9 & 2006 & 2004 & 106 & 91 \\
0 & 60 & 68 & 8 & 2004 & 2004 & 96 & 81 \\
1 & 81 & 88 & 7 & 2020 & 2008 & 181 & 14 \\
1 & 81 & 88 & 7 & 2014 & 2020 & 206 & 93 \\
1 & 78 & 84 & 6 & 2012 & 2014 & 215 & 214 \\
1 & 79 & 84 & 5 & 2005 & 2005 & 166 & 86 \\
1 & 79 & 84 & 5 & 2011 & 2010 & 86 & 81 \\
1 & 81 & 86 & 5 & 2018 & 2016 & 91 & 91 \\
1 & 82 & 87 & 5 & 2010 & 2017 & 181 & 131 \\
1 & 76 & 80 & 4 & 2005 & 2006 & 212 & 213 \\
1 & 79 & 83 & 4 & 2009 & 2007 & 105 & 91 \\
1 & 79 & 83 & 4 & 2004 & 2007 & 101 & 91 \\
2 & 87 & 96 & 9 & 2009 & 2007 & 26 & 27 \\
2 & 88 & 95 & 7 & 2012 & 2012 & 55 & 158 \\
2 &	87 & 93 & 6 & 2019 & 2019 &	26 & 27 \\
2 &	87 & 93 & 6 & 2019 & 2019 &	32 & 27 \\
2 & 84 & 89 & 5 & 2012 & 2014 & 213 & 213 \\
2 & 81 & 85 & 4 & 2004 & 2005 & 212 & 213 \\
2 & 82 & 85 & 3 & 2011 & 2009 & 174 & 172 \\
2 & 82 & 85 & 3 & 2018 & 2019 & 175 & 175 \\
2 & 86 & 89 & 3 & 2011 & 2011 & 84 & 82 \\
2 & 89 & 92 & 3 & 2015 & 2016 & 162 & 162 \\
\bottomrule
\end{tabular}}
\caption{Top $10$ most persistent features in each dimension for the group `Comunidad de Madrid'}
\end{table}

\begin{table}\centering
\resizebox{!}{0.4\textwidth}{\begin{tabular}{lrrrrrrr}
dim&birth&death&persistence&year of birth&year of death&neighbourhood of birth&neighbourhood of death\\
\midrule
0 & 53 & 100 & 47 & 2020 &  & 141 & \\
0 & 58 & 78 & 20 & 2021 & 2020 & 81 & 144 \\
0 & 61 & 77 & 16 & 2020 & 2023 & 212 & 42 \\
0 & 62 & 76 & 14 & 2018 & 2023 & 27 & 93 \\
0 & 66 & 80 & 14 & 2016 & 2019 & 141 & 141 \\
0 & 67 & 81 & 14 & 2009 & 2007 & 141 & 145 \\
0 & 71 & 83 & 12 & 2018 & 2019 & 194 & 194 \\
0 & 68 & 79 & 11 & 2020 & 2020 & 194 & 191 \\
0 & 68 & 77 & 9 & 2013 & 2016 & 27 & 41 \\
0 & 72 & 80 & 8 & 2004 & 2005 & 158 & 158 \\
1 & 79 & 84 & 5 & 2012 & 2017 & 85 & 84 \\
1 & 76 & 80 & 4 & 2020 & 2021 & 16 & 11 \\
1 & 78 & 82 & 4 & 2009 & 2010 & 76 & 93 \\
1 & 79 & 83 & 4 & 2023 & 2019 & 88 & 86 \\
1 & 79 & 83 & 4 & 2020 & 2020 & 191 & 193 \\
1 & 79 & 83 & 4 & 2023 & 2023 & 191 & 193 \\
1 & 79 & 83 & 4 & 2015 & 2017 & 55 & 157 \\
1 & 79 & 83 & 4 & 2008 & 2008 & 11 & 102 \\
1 & 79 & 83 & 4 & 2005 & 2006 & 93 & 93 \\
1 & 79 & 83 & 4 & 2023 & 2021 & 205 & 143 \\
2 & 83 & 95 & 12 & 2014 & 2015 & 25 & 27 \\
2 & 85 & 95 & 10 & 2009 & 2009 & 24 & 27 \\
2 & 80 & 89 & 9 & 2019 & 2019 & 26 & 27 \\
2 & 86 & 95 & 9 & 2015 & 2015 & 131 & 27 \\
2 &	86 & 95 & 9 & 2011 & 2009 & 132 & 27 \\
2 &	83 & 89 & 6 & 2017 & 2009 & 166 & 158 \\
2 &	83 & 89 & 6 & 2019 & 2019 & 33 & 27 \\
2 & 84 & 90 & 6 & 2011 & 2012 & 27 & 27 \\
2 & 83 & 88 & 5 & 2023 & 2019 & 87 & 82 \\
2 & 85 & 90 & 5 & 2017 & 2018 & 97 & 95 \\
\bottomrule
\end{tabular}}
\caption{Top $10$ most persistent features in each dimension for the group `outside'}
\end{table}

\clearpage

\begin{table}[t]
\centering
\begin{tabular}{ccc}
\toprule
neighbourhood & year & prevalence \\
\midrule
27 & 2018 & 6 \\
27 & 2019 & 2 \\
\bottomrule
\end{tabular}
\caption{List of the neighbourhood-years enclosed at birth-time in the most persistent $H_2$ feature for the group `stay', together with their prevalence.}
\end{table}

\begin{table}[h]
\centering
\begin{tabular}{ccc}
\toprule
neighbourhood & year & prevalence \\
\midrule
25 & 2016 & 10 \\
27 & 2013 & 5 \\
27 & 2014 & 7 \\
27 & 2015 & 3 \\
27 & 2016 & 6 \\
\bottomrule
\end{tabular}
\caption{List of the neighbourhood-years enclosed at birth-time in the second most persistent $H_2$ feature for the group `stay', together with their prevalence.}
\end{table}

\begin{table}[h]
\centering
\begin{tabular}{ccc}
\toprule
neighbourhood & year & prevalence \\
\midrule
27 & 2013 & 5 \\
27 & 2014 & 7 \\
27 & 2015 & 3 \\
27 & 2016 & 6 \\
35 & 2013 & 8 \\
35 & 2015 & 6 \\
\bottomrule
\end{tabular}
\caption{List of the neighbourhood-years enclosed at birth-time in the third most persistent $H_2$ feature for the group `stay', together with their prevalence.}
\end{table}

\newpage

\begin{table}[t]
\centering
\begin{tabular}{ccc}
\toprule
neighbourhood & year & prevalence \\
\midrule
27 & 2018 & 35 \\
\bottomrule
\end{tabular}

\caption{List of the neighbourhood-years enclosed at birth-time in the  most persistent $H_2$ feature for the group `city', together with their prevalence.}
\end{table}

\begin{table}[h]
\centering
\begin{tabular}{ccc}
\toprule
neighbourhood & year & prevalence \\
\midrule
27 & 2020 & 35 \\
27 & 2021 & 42 \\
\bottomrule
\end{tabular}

\caption{List of the neighbourhood-years enclosed at birth-time in the second most persistent $H_2$ feature for the group `city', together with their prevalence.}
\end{table}

\begin{table}[h]
\centering
\begin{tabular}{ccc}
\toprule
neighbourhood & year & prevalence \\
\midrule
141 & 2020 & 35 \\
\bottomrule
\end{tabular}

\caption{List of the neighbourhood-years enclosed at birth-time in the third most persistent $H_2$ feature for the group `city', together with their prevalence.}
\end{table}

\clearpage

\begin{table}[t]
\centering
\begin{tabular}{ccc}
\toprule
neighbourhood & year & prevalence \\
\midrule
27 & 2006 & 5 \\
27 & 2007 & 4 \\
27 & 2008 & 9 \\
27 & 2009 & 10 \\
\bottomrule
\end{tabular}
\caption{List of the neighbourhood-years enclosed at birth-time in the  most persistent $H_2$ feature for the group `C. Madrid', together with their prevalence.}
\end{table}

\begin{table}[h]
\centering
\begin{tabular}{ccc}
\toprule
neighbourhood & year & prevalence \\
\midrule
158 & 2012 & 5 \\
\bottomrule
\end{tabular}
\caption{List of the neighbourhood-years enclosed at birth-time in the second most persistent $H_2$ feature for the group `C. Madrid', together with their prevalence.}
\end{table}

\begin{table}[h]
\centering
\begin{tabular}{ccc}
\toprule
neighbourhood & year & prevalence \\
\midrule
27 & 2019 & 7 \\
\bottomrule
\end{tabular}

\caption{List of the neighbourhood-years enclosed at birth-time in the third most persistent $H_2$ feature for the group `C. Madrid', together with their prevalence.}
\end{table}

\clearpage

\begin{table}[t]
\centering
\begin{tabular}{ccc}
\toprule
neighbourhood & year & prevalence \\
\midrule
27 & 2015 & 5 \\
27 & 2016 & 16 \\
27 & 2017 & 12 \\
\bottomrule
\end{tabular}
\caption{List of the neighbourhood-years enclosed at birth-time in the  most persistent $H_2$ feature for the group `outside', together with their prevalence.}
\end{table}

\begin{table}[h]
\centering
\begin{tabular}{ccc}
\toprule
neighbourhood & year & prevalence \\
\midrule
27 & 2007 & 11 \\
27 & 2008 & 8 \\
27 & 2009 & 5 \\
27 & 2010 & 10 \\
\bottomrule
\end{tabular}
\caption{List of the neighbourhood-years enclosed at birth-time in the second most persistent $H_2$ feature for the group `outside', together with their prevalence.}
\end{table}

\begin{table}[h]
\centering
\begin{tabular}{ccc}
\toprule
neighbourhood & year & prevalence \\
\midrule
27 & 2019 & 11 \\
\bottomrule
\end{tabular}

\caption{List of the neighbourhood-years enclosed at birth-time in the third most persistent $H_2$ feature for the group `outside', together with their prevalence.}
\end{table}

\clearpage

\section{Flowchart}

\begin{figure}[h]
    \centering
    \begin{tikzpicture}[
  font=\sffamily\footnotesize, ink,
  >={Stealth[length=2.2mm]},
  pbox/.style={draw=ink, rounded corners=3pt, fill=white, line width=0.6pt,
               text width=4.3cm, minimum height=2.45cm, align=flush left,
               inner sep=5pt, anchor=north west},
  wide/.style={pbox, text width=4.9cm},
  num/.style={circle, fill=ink, text=white, font=\sffamily\bfseries\scriptsize,
              inner sep=1.2pt, minimum size=4.2mm},
  tag/.style={font=\sffamily\scriptsize\itshape, text=ink!60, anchor=north east},
  flow/.style={->, line width=0.9pt, ink},
  phase/.style={font=\sffamily\bfseries\footnotesize, text=ink!70},
]

\node[pbox] (s1) at (0,0) {%
  \textbf{Raw data}\\[2pt]
  Yearly origin--destination matrices $M^{t}$, $t=2004,\dots,2023$.
  Rows: 131 origin neighbourhoods; columns: destinations (131
  neighbourhoods, other towns in the region, other regions, abroad).
  Plus the shapefile.};
\node[pbox] (s2) at ($(s1.north east)+(7mm,0)$) {%
  \textbf{Harmonise boundaries}\\[2pt]
  Neighbourhoods created in 2017 (183, 193, 194) and the re-coded 192
  take their parent area's values for 2004--2016, so the map has
  no artificial holes.};
\node[wide] (s3) at ($(s2.north east)+(7mm,0)$) {%
  \textbf{Collapse destinations into 4 groups}\\[2pt]
  $g\in\{\text{stay},\ \text{city},\ \text{C.\,Madrid},\ \text{outside}\}$:
  $M^{t}\mapsto R^{t}\in\mathbb{R}^{131\times4}$, then row-normalise\\[3pt]
  \hfil$p_g(n,t)=100\,R^{t}_{ng}\big/\textstyle\sum_{h}R^{t}_{nh}$};

\foreach \s/\n/\t in {s1/1/{\S4.1, SI \S A.1}, s2/2/{SI \S A.2}, s3/3/{\S2.1, \S4.4}}{
  \node[num] at (\s.north west) {\n};
  \node[tag] at ($(\s.south east)+(0,-1pt)$) {\t};
}
\draw[flow] (s1.east |- s1.center) -- (s2.west |- s1.center);
\draw[flow] (s2.east |- s1.center) -- (s3.west |- s1.center);

\node[pbox] (s4) at ($(s1.south west)+(0,-14mm)$) {%
  \textbf{Rasterise and stack}\\[2pt]
  $100\times100$ grid over the map; pixel $(x,y)$ gets its neighbourhood
  $\nu(x,y)$. Stacking the 20 years gives a $100\times100\times20$ image\\[3pt]
  \hfil$I_g(x,y,z)=p_g\big(\nu(x,y),z\big)$};
\node[pbox] (s5) at ($(s4.north east)+(7mm,0)$) {%
  \textbf{Superlevel filtration}\\[2pt]
  $K_g(r)=\{\text{voxels with } I_g\ge r\}$, $r=100\to0$.
  T-construction: each voxel enters at its own value; a face enters with
  its first adjacent voxel.};
\node[wide] (s6) at ($(s5.north east)+(7mm,0)$) {%
  \textbf{Persistent homology}\\[2pt]
  Cubical Ripser gives the diagrams $\mathrm{Dgm}_k(I_g)$, $k=0,1,2$:
  every feature $(b,d)$ together with its birth and death voxels $(x,y,z)$.};

\node[wide] (s7) at ($(s6.south west)+(0,-9mm)$) {%
  \textbf{Post-process}\\[2pt]
  Map voxels back to (neighbourhood, year); recover the voxels enclosed by
  each $H_2$ class at birth; drop duplicate features; keep persistence $>2$ and rank the top~10.};
\node[pbox, anchor=north east, text width=9.27cm] (s8) at ($(s7.north west)+(-7mm,0)$) {%
  \textbf{Read the features as displacement evidence}\\[3pt]
  \begin{tabular}{@{}p{1.15cm}@{}>{\raggedright\arraybackslash}p{8.1cm}@{}}
    $H_0$ & an enclave where group $g$ is prevalent (birth $=$ peak share)\\[1pt]
    $H_1,H_2$ & a set of adjacent neighbourhood-years where $g$ is
    \emph{under}-represented compared to its surroundings; for $H_2$ this set
    is bounded in both space and time
  \end{tabular}\\[4pt]
  {\color{cstay}\rule{6pt}{6pt}}\,{\color{ccity}\rule{6pt}{6pt}}\;
  holes in \emph{stay} or \emph{city}\;$\Rightarrow$\;\textbf{displacement}\\[1pt]
  {\color{ccm}\rule{6pt}{6pt}}\,{\color{cout}\rule{6pt}{6pt}}\;
  holes in \emph{C.\,Madrid} or \emph{outside}\;$\Rightarrow$\;\textbf{eased pressure}};

\foreach \s/\n/\t in {s4/4/{\S4.4, Fig.~4}, s5/5/{\S4.4, SI \S B},
                      s6/6/{\S4.4, Figs.~6--7}, s7/7/{\S4.6, SI tables},
                      s8/8/{\S4.5, \S4.7}}{
  \node[num] at (\s.north west) {\n};
  \node[tag] at ($(\s.south east)+(0,-1pt)$) {\t};
}
\draw[flow] (s4.east |- s4.center) -- (s5.west |- s4.center);
\draw[flow] (s5.east |- s4.center) -- (s6.west |- s4.center);
\draw[flow] (s6.south) -- (s7.north);
\draw[flow] (s7.west |- s8.center) -- (s8.east |- s8.center);

\begin{scope}[shift={($(s4.north east)+(-8mm,-3.6mm)$)}, line width=0.35pt]
  \foreach \k/\shade in {0/55,1/38,2/22,3/8}
    \draw[fill=ink!\shade, draw=ink!60]
      (0,\k*0.75mm) -- ++(4mm,0) -- ++(1.5mm,1.1mm) -- ++(-4mm,0) -- cycle;
\end{scope}
\begin{scope}[shift={($(s6.north east)+(-9mm,-2.4mm)$)}, line width=1.1pt]
  \draw[ink!55] (0,0) -- (6.5mm,0);
  \draw[ink!55] (1mm,-0.8mm) -- (3mm,-0.8mm);
  \draw[ink!55] (2mm,-1.6mm) -- (7mm,-1.6mm);
  \draw[ink!55] (0.5mm,-2.4mm) -- (2mm,-2.4mm);
\end{scope}

\coordinate (pad1) at ($(s3.south)+(0,-4mm)$);
\coordinate (pad2) at ($(s8.south)+(0,-4.5mm)$);
\begin{scope}[on background layer]
  \node[fill=soft, rounded corners=6pt, inner sep=3.5mm, fit=(s1)(s2)(s3)(pad1)] (bg1) {};
  \node[draw=rule, dashed, rounded corners=6pt, inner sep=3.5mm,
        fit=(s4)(s5)(s6)(s7)(s8)(pad2)] (bg2) {};
\end{scope}
\node[phase, anchor=south west] at ($(bg1.north west)+(1mm,0.3mm)$)
  {A.\ Data preparation (done once)};
\node[phase, anchor=north west] at ($(bg2.south west)+(1mm,-0.6mm)$)
  {B.\ Topological analysis (steps 4--8 run separately for each group $g$)};

\path let \p1=(bg1.south), \p2=(bg2.north) in coordinate (midY) at (0,{(\y1+\y2)/2});
\foreach \c/\k in {cstay/0, ccity/1, ccm/2, cout/3}{
  \draw[\c, line width=1.5pt, rounded corners=5pt]
    ($(s3.south)+(-4.5mm+\k*3mm,0)$) |- ($(midY -| s4.north)+(-4.5mm+\k*3mm,1.8mm-\k*1.2mm)$)
    -- ($(s4.north)+(-4.5mm+\k*3mm,0)$);
}
\node[font=\sffamily\scriptsize, fill=white, inner sep=1.5pt, rounded corners=2pt]
  at ($(midY -| s5.north)$) {one 3D image per group:\enspace
  {\color{cstay}\rule{5pt}{5pt}}\,stay\enspace {\color{ccity}\rule{5pt}{5pt}}\,city\enspace
  {\color{ccm}\rule{5pt}{5pt}}\,C.\,Madrid\enspace {\color{cout}\rule{5pt}{5pt}}\,outside};
\end{tikzpicture}
    \caption{Overview of the pipeline. Panel A runs once on the raw data; panel B runs
separately for each of the four groups (one colour per group), producing one 3D
image, one filtration and one set of persistence diagrams per group. The grey tag
under each step gives the section, figure or Supporting Information (SI) section
where that step is described in detail.}
\label{fig:pipeline}
\end{figure}

\clearpage

\section{$H_1$ holes}

\begin{figure}[ht]
\centering
\includegraphics[width=\linewidth]{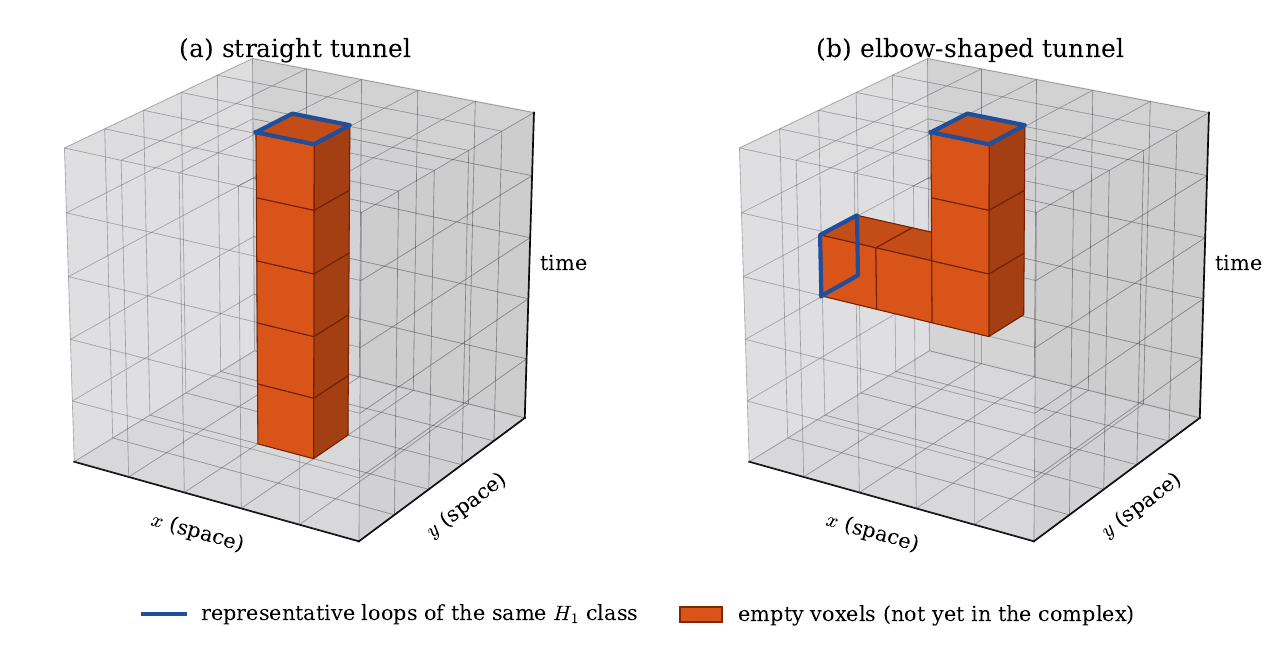}
\caption{Two $1$--dimensional holes ($H_1$ features) in a $5\times5\times5$ toy cubical complex, with space on the horizontal axes and time on the vertical one; the lateral faces of the block represent the city border. Orange voxels are not yet in the complex at the current filtration value. (a) Straight tunnel: a single neighbourhood where the group is not prevalent over the whole period, as cell A in Figure~\ref{fig:toy_PH}. (b) Elbow-shaped tunnel: a neighbourhood where the group is not prevalent during the last years, which in a given year extends horizontally across adjacent neighbourhoods up to the city border. In both cases the empty region is connected to the exterior of the complex, so there is no canonical set of voxels enclosed by the hole. The blue loops lie on the boundary of the complex and surround the openings of the tunnel; in (b), the two loops lie on different faces of the complex but represent the same $H_1$ class.}
\label{fig:SI_tunnels}
\end{figure}

\end{document}